\catcode`\@=11
\newif\if@fewtab\@fewtabtrue
{\count255=\time\divide\count255 by 60
\xdef\hourmin{\number\count255}
\multiply\count255 by-60\advance\count255 by\time
\xdef\hourmin{\hourmin:\ifnum\count255<10 0\fi\the\count255}}
\def\ps@draft{\let\@mkboth\@gobbletwo
    \def\@oddhead{}
    \def\@oddfoot
       {\hbox to 7 cm{$\scriptstyle Draft\ version:\ \draftdate$
       \hfil}\hskip -7cm\hfil\rm\thepage \hfil}
    \def\@evenhead{}\let\@evenfoot\@oddfoot}

\def\ceqno{\global\@fewtabfalse
    \ifcase\@eqcnt \def\@tempa{& & &}\or \def\@tempa{& &}
      \or \def\@tempa{&}
      \or\def\@tempa{}\fi\@tempa
{\rm(\theequation)}}

\def\aeqno#1{\global\@fewtabfalse
    \ifcase\@eqcnt \def\@tempa{& & &}\or \def\@tempa{& &}
      \or \def\@tempa{&}
      \or\def\@tempa{}\fi\@tempa
{\rm(\theequation,#1)}}

\def\label#1{\ifnum\draftcontrol=1
 \global\def\draftnote{$\scriptstyle #1$}\fi
 \@bsphack\if@filesw {\let\thepage\relax
   \def\protect{\noexpand\noexpand\noexpand}%
\xdef\@gtempa{\write\@auxout{\string
      \newlabel{#1}{{\@currentlabel}{\thepage}}}}}\@gtempa
   \if@nobreak \ifvmode\nobreak\fi\fi\fi
  \@esphack}

\def\alabel#1#2{\label{#1}\global\@fewtabfalse
    \ifcase\@eqcnt \def\@tempa{& & &}\or \def\@tempa{& &}
      \or \def\@tempa{&}
      \or\def\@tempa{}\fi\@tempa
{\hbox to 3cm{\phantom{\rm(\theequation,#2)}
\draftnote \hfil}\hskip -3cm {\rm(\theequation,#2)}}}

\def\clabel#1{\label{#1}\global\@fewtabfalse
    \ifcase\@eqcnt \def\@tempa{& & &}\or \def\@tempa{& &}
      \or \def\@tempa{&}
      \or\def\@tempa{}\fi\@tempa
{\hbox to 3cm{\phantom{\rm(\theequation)}
\draftnote \hfil}\hskip -3cm{\rm(\theequation)}}}

\def\eqnarray{\def\draftnote{{}}\global\@fewtabtrue
\stepcounter{equation}\let\@currentlabel=\theequation
\global\@eqnswtrue
\global\@eqcnt\z@\tabskip\@centering\let\\=\@eqncr
$$\halign to \displaywidth\bgroup\@eqnsel\hskip\@centering\@eqcnt\z@
  $\displaystyle\tabskip\z@{##}$&\global\@eqcnt\@ne
  \hskip 1\arraycolsep \hfil${##}$\hfil
  &\global\@eqcnt\tw@ \hskip 1\arraycolsep
$\displaystyle\tabskip\z@{##}$
\hfil  \tabskip\@centering&\global\@eqcnt\thr@@\llap{##}\tabskip\z@
\cr}

\def\endeqnarray{\@@eqncr\egroup
      \global\advance\c@equation\m@ne$$\global\@ignoretrue}

\def\@eqnnum{\hbox to 3cm{\phantom{\rm(\theequation)} \draftnote
                         \hfil}\hskip -3cm {\rm(\theequation)}}

\def\@@eqncr{\let\@tempa\relax
    \ifcase\@eqcnt \def\@tempa{& & &}\or \def\@tempa{& &}
      \or \def\@tempa{&}
      \or\def\@tempa{}
\fi\@tempa
\if@eqnsw
\if@fewtab\@eqnnum\fi
\stepcounter{equation}\fi\global
\@eqnswtrue\global\@eqcnt\z@\global\@fewtabtrue\cr}

\def\draftcite#1{\ifnum\draftcontrol=1#1\else{}\fi}

\def\@lbibitem[#1]#2{\item{}\hskip -3cm \hbox to 2cm
{\hfil$\scriptstyle\draftcite{#2}$}\hskip
1cm[\@biblabel{#1}]\if@filesw
     {\def\protect##1{\string ##1\space}\immediate
      \write\@auxout{\string\bibcite{#2}{#1}}}\fi\ignorespaces}

\def\@bibitem#1{\item\hskip -3cm \hbox to 2cm
{\hfil $\scriptstyle\draftcite{#1}$}\hskip 1cm
\if@filesw \immediate\write\@auxout
       {\string\bibcite{#1}{\the\value{\@listctr}}}\fi\ignorespaces}

\def\nsection#1{\section{#1}\setcounter{equation}{0}}

\def\CA{{\cal A}}       \def\CB{{\cal B}}       
\def\CD{{\cal D}}              \def\CF{{\cal F}}
\def\CG{{\cal G}}              \def\CI{{\cal J}}
              \def\CL{{\cal L}}
              \def\CO{{\cal O}}
              
       \def\CT{{\cal T}}

\newcommand{\NR}{{{\bf R}}}
\newcommand{\NP}{{{\bf P}}}
\newcommand{\NC}{{{\bf C}}}
\newcommand{\NT}{{{\bf T}}}
\newcommand{\NZ}{{{\bf Z}}}

\newcommand{\NQ}{{{\bf Q}}}

\newcommand{\hslash}{{h\hspace{-0.23cm}^-}}

\def\qq{ \begin{eqnarray} }
\def\qqq{ \end{eqnarray} }
\def\non{ \nonumber }
\newcommand{\no}{\noindent}
\newcommand{\vs}{\vspace}

\newcommand{\da}{\partial}
\newcommand{\ee}{{\rm e}}

\newcommand{\hf}{{_1\over^2}}
\newcommand{\ha}{{1\over 2}}

\newcommand{\m}{{\hspace{0.025cm}}}

\def\draftdate{\number\month/\number\day/\number\year\ \ \ \hourmin }

\global\def\draftcontrol{0}
\catcode`\@=12
\documentstyle[12pt]{article}

\def\theequation{{\thesection.\arabic{equation}}}

\begin{document}
\begin{center}

{\Large{\bf{KAM Theorem and Quantum Field Theory}}}

\vs{ 1cm}

{\large{Jean Bricmont}\footnote{Partially supported by
 EC grant CHRX-CT93-0411}}

\vs{ 0.2cm}

UCL, FYMA, 2 chemin du Cyclotron,\\ 
B-1348  Louvain-la-Neuve, Belgium

\vs{0.5cm}

{\large{Krzysztof Gaw\c{e}dzki}}
\vs{ 0.2cm}

CNRS, IHES, 35 route de Chartres,\\ 
91440 Bures-sur-Yvette, France

\vs{0.5cm}

{\large{Antti Kupiainen}}\footnote{Partially supported by
NSF grant DMS-9501045 and EC grant CHRX-CT93-0411}

\vs{ 0.2cm}

Department of Mathematics,
Helsinki University,\\
P.O. Box 4, 00014 Helsinki, Finland
\end{center}
\date{ }

\vskip 0.3cm
\vskip 1.3 cm

\begin{abstract}
\vskip 0.3cm

\noindent  We give a new proof  
of the KAM theorem for analytic Hamiltonians. 
The proof is inspired by a quantum field theory 
formulation of the problem and is based on a renormalization group 
argument treating the small denominators inductively scale by scale. 
The crucial cancellations of resonances are shown to follow from 
the  Ward identities expressing the translation invariance 
of the corresponding field theory.
\end{abstract}
\vs{ 1.6cm}

\nsection{Introduction}
\vskip 0.2cm

Consider the Hamiltonian 
\qq
H(I,\phi)\,=\,\omega \cdot I +  \hf\m I\cdot\mu I+\lambda\, U(\phi,I)
\label{H}
\qqq
with $\phi\in\NR^d/(2\pi\NZ^d)\equiv\NT^d$,  $I\in\NR^d$, 
$\omega\in\NR^d$ with the components
$\omega_i$ independent over $\NZ^d$ and $\mu$ a real symmetric
 $d\times d$ matrix.
It generates the Hamiltonian flow given by the equations of motion
\qq
\dot\phi=\omega +\mu  I+\lambda\,\partial_I U\, ,\;\;\dot I=
-\lambda\m\partial_\phi U\m.
\label{eqs}
\qqq
For the parameter $\lambda=0$ and the initial condition 
$(\phi_0,0)$, the flow $(\phi_0+\omega t,0)$
is quasiperiodic and spans a $d$-dimensional torus in 
$\NT^d\times\NR^d$. KAM-theorem deals with the question
under what conditions such quasiperiodic solutions
persist as the parameter $\lambda$ is turned on.
\vskip 0.2cm

Consider a quasiperiodic solution in the form
$$
(\phi(t),I(t))=(\phi_0+\omega t+\Theta(\phi_0+\omega t),
\, J(\phi_0+\omega t))\m.
$$
Eqs.\,\,(\ref{eqs}) require that $Z=(\Theta,J):\NT^d
\rightarrow \NR^d\times\NR^d$ satisfies the relation
\qq
\CD Z(\phi)=-\lambda\,\partial U(\phi+\Theta(\phi),
\, J(\phi))\m,
\label{Z}
\qqq
where $\partial=(\partial_\phi ,\partial_I)$ and
\qq
\CD=\left(\matrix{0 & \omega\cdot\partial_\phi \cr 
-\omega\cdot\partial_\phi & \mu}\right).
\label{G}
\qqq
Note that if $Z$ is a solution of Eq.\,\,(\ref{Z}) 
then so is $Z_\beta$
for $\beta\in  \NR^d$ and
\qq
Z_\beta(\phi)\,=\, Z(\phi-\beta)-(\beta,0)\m.
\label{Ztr}
\qqq
Eq.\,\,(\ref{Z}) is a fixed point problem for the function $Z$ 
of a difficult type: the straightforward linearization
$\CD+\lambda\da\da U(\phi,0)$ is not invertible
for any interesting $U$ (see e.g. \cite{E}). Also, 
one can expect to have a solution only for
sufficiently irrational $\omega\in\NR^d$, e.g.
satisfying a Diophantine condition
\qq
|\omega\cdot q|\, >\, a\m |q|^{-\nu}\quad{\rm for}\quad q\in \NZ^d,
\ q\neq 0
\label{Dio}
\qqq
with some $q$-independent $a,\nu>0$. There
have been traditionally two approaches to the problem:

\vs{2mm}

\no 1. The KAM approach. (\ref{Z}) is solved by a
Newton method that constructs a sequence of
symplectic changes of coordinates defined on
shrinking domains that, in the limit, transform
the problem to the $\lambda=0$ case \cite{A1,A2,Ko,Mo1}.

\vs{2mm}

\no 2. Perturbation theory. For $U$ analytic
(see below) one can attempt to solve (\ref{Z})
by iteration. This leads to a power series in
$\lambda$, the Lindstedt series: $Z=\sum_n Z_n\lambda^n$.
Each $Z_n$ is given as a sum of several
terms (see Sect.\,\,9), some of which are very
large, proportional to $(n!)^a$ with $a>0$,
due to piling up of ``small denominators'' $(\omega\cdot q)$ 
from the momentum space representation of operator $\CD^{-1}$. 
However, the KAM method also  yields  the analyticity
of $Z$ in $\lambda$ \cite{Mo2}. Thus the Lindstedt series
must converge. To see this directly turned out to be
rather hard and was finally done by
Eliasson \cite{E} who, by regrouping terms, 
was able to produce an absolutely convergent 
series that gives the quasiperiodic solution. 
Subsequently Eliasson's work was simplified and extended
by Gallavotti \cite{G1,G2, GalPar}, by
Chierchia and Falcolini \cite{CF1,CF2} and by
 Bonetto, Gentile, Mastropietro 
\cite{GGM,GM1,GM2,GM3,BGGM1,BGGM2}.

\vs{4mm}

In the present paper we shall develop a new iterative
scheme to solve Eq.\,\,(\ref{Z}). It is based 
on a direct application of the renormalization group (RG) idea 
of quantum field theory (QFT) to the problem.
The idea is to split the operator $\CD$ (or rather its
inverse, see Sect.\,\,2) into a small denominator and 
large denominator part, where small and large
are defined with respect to a scale of order unity. The next step
is to solve the large denominator problem which results
in a new effective equation of the type (\ref{Z}) for
the small denominator part, with a new right hand side.
The procedure is iterated, with the scale separating
small and large at the $n^{\rm th}$ step equal to
$\eta^{n}$ for some fixed $\eta<1$. As a result we get 
a sequence of effective problems that converge to a trivial 
one as $n\rightarrow \infty$. A generic step is solved 
by a simple application of the Banach Fixed Point Theorem
in a big space of functionals of $Z$
representing the right hand side of Eq.\,\,(\ref{Z}) 
in the $n^{\rm th}$ iteration step. 

\vs{2mm}

Our iteration can be viewed as an iterative resummation 
of the Lindstedt series, as will be discussed in Sect.\,\,9. 
This iterative approach trivializes the rather formidable
combinatorics of the small denominators.
The functional formulation in terms
of effective problems removes also the mystery
behind the subtle cancellations in the Lindstedt series:
they turn out to be an easy consequence
of a symmetry in the problem as formulated in terms
of the so called Ward identities of QFT.
The QFT analogy of the problem (\ref{Z}) has been
forcefully emphasized by Gallavotti {\it et al.} \cite{GalPar,GGM}. The
proof of Eliasson's theorem by these authors was based 
on a separation into scales of the graphical expressions
entering the Lindstedt series and was a direct
inspiration for the present work.
\vs{2mm}

An important part of the standard RG theory is
an approximate scale invariance of the problem
that is exhibited and exploited by the RG
method. The KAM problem also is expected
to have this aspect: as the coupling $\lambda$
is increased the solution with a given
$\omega$ eventually ceases to exist. For
suitable ``scale invariant'' $\omega$ (e.g.
in $d=2$ for $\omega=(1,\gamma)$ with $\gamma$
a ``noble'' irrational) the solution at the critical 
$\lambda$ is expected to exhibit a power law decay
of Fourier coefficients and periodic
orbits converging to it have peculiar
``universal'' scaling properties \cite{kad,she}.
We hope that the present approach will
shed some light on these problems in the future.

\vskip 0.4cm

While the main goal of this paper
 is to develop a new method, we use
it to reprove the following (classical) result:

\vs{4mm}

\no{\bf Theorem 1}. {\ \it Let $U$ be real analytic in
$\phi$ and analytic in $I$ in a neighborhood
of $I=0$. Assume that  $\omega$ satisfies condition}
(\ref{Dio}). {\it Then Eq}.\,\,(\ref{Z}) {\it has a solution 
which is analytic in $\lambda$ and real analytic in $\phi$ 
provided that either}

\vs{2mm}

\no (a) (the non-isochronous case) \ $\mu$ {\it is an invertible
matrix and $\vert\lambda\vert$ is small enough (in a $\mu$-dependent 
way)}.
\vs{2mm}

\no (b) (the isochronous case) \ $\mu =0$, 
$\,\int_{_{\NT^d}}\hspace{-0.05cm}
\partial_{I} U(\phi,0)\, d\phi =0$, {\it\,the
$d\times d$ matrix with elements
$\int_{\NT^d}\partial_{I_k}\partial_{I_l} U(\phi,0)\, d\phi$, 
$\m k,l = 1,\dots, d\m,$ \,is invertible and 
$\vert\lambda\vert$ is  small enough.}
\vs{2mm}

\noindent{\it The above solutions are unique up to translations 
(\ref{Ztr}).}
\vs{4mm}

\no{\bf Remark}. \ Actually, we show that the solution is an
analytic function not only of $\lambda$, but of the potential
$U$, when the latter belongs to a small ball in a Banach
space of analytic functions (see Sect.\,\,3 for
the introduction of such spaces). This allows us to consider 
more general Hamiltonians of the form
\qq
H(I,\phi)\,=\,H_0 (I) +\, U(\phi,I).
\nonumber
\qqq
with $H_0$ and $U$ analytic and $U$ small. Indeed, we may 
expand $H_0$ around
$I_0$ s.t. $\partial_I H_0 (I_0) = \omega$, with $\omega$
satisfying condition (\ref{Dio}): 
$$H_0(I)=H(I_0) +\omega \cdot (I-I_0) +\hf\,(I-I_0)\cdot
\mu(I-I_0)+{\tilde H}_0 (I)$$ 
and define ${\tilde U}=U +{\tilde H}_0$  
so as to include in it all the terms of order higher than
two in the expansion of $H_0$. Replacing $I-I_0$ by $I$, 
we may apply Theorem 1 provided that ${\tilde U}$ satisfies
the corresponding hypotheses. Also, more general cases where 
$\mu$ is a degenerate matrix can be treated.
\vs{4mm}

The organization of the paper is as follows. In Sect.\,\,2
we explain the RG formalism. In  Sect.\,\,3, we introduce
spaces of analytic functions on Banach spaces; such spaces
will be used to solve our RG equations.  In
Sect.\,\,4, we state the main inductive estimates which
are proved in Sect.\,\,6 after an interlude on
the Ward identities in Sect.\,\,5. Theorem 1 is proved
then in Sect.\,\,7. Sect.\,\,8 explains the connection
of our formalism to QFT for those familiar with the
latter. We should emphasize that the QFT is solely 
a source of intuition, the simple RG formalism of
Sect.\,\,2 is independent of it. Finally, in
Sect.\,\,9, the connection with the  Lindstedt series is
explained.

\nsection{Renormalization group scheme}
\vs{2mm}

In this section we explain the iterative RG scheme without spelling
out the technical assumptions that are needed to carry it out.
We refer the reader to Sect.\,\,9 for a graphical representation
of the main quantities introduced here.
\vskip 0.4cm

We shall work with Fourier transforms, denoting by  lower case 
letters the Fourier transforms of functions of $\phi$,
the latter being  denoted
by capital letters:
\qq
F(\phi)=\sum\limits_{q\in{\NZ}^d}\ee^{-i\m q\cdot\phi}
\, f(q)\m,\quad\ {\rm where}\quad\ f(q)=\int_{\NT^d}
\ee^{i\m q\cdot\phi}\, F(\phi)\, d\phi
\nonumber
\qqq
with  $d\phi$ standing for the normalized Lebesgue measure
on $\NT^d$. 
\vskip 0.4cm

Note first that we may use the translations
(\ref{Ztr}) to limit our search for the solution of Eq.\,\,(\ref{Z})
to the subspace of $\Theta$ with zero average, i.e. 
with $\theta(0)=0$ in the Fourier language.
It will be convenient to separate the constant mode of $J$
explicitly by writing $Z=X+(0,\zeta)$ where $X$ has zero average.  
Let us define
\qq
W_0(\phi;X,\zeta)=
\lambda\,\partial U((\phi,\zeta)+X(\phi))\m.
\label{W1}
\qqq
Denote by $G_0$ the operator $-\CD^{-1}$ 
acting on $\NR^{2d}$-valued functions on $\NT^d$ 
with zero average. In terms of the Fourier transforms, 
\qq
(G_0\m x)(q)=\left(\matrix{\mu(\omega\cdot q)^{-2} & 
i(\omega\cdot q)^{-1} \cr 
-i(\omega\cdot q)^{-1} & 0}\right)x(q) .
\label{G1}
\qqq
for $q\neq 0$ and $(G_0\m x)(0)=0$.
Writing Eq.\,\,(\ref{Z}) separately for the averages (i.e. $q=0$) 
and the rest, we may rewrite it as the fixed point 
equations
\qq
X &= &G_0\m P\m W_0(X,\zeta)\m,
\label{fp1}\\
(0,\mu\zeta)&=&-\int_{\NT^d} W_0(\phi;X,\zeta)\, d\phi\m,\label{fp12}
\qqq
where $P$ projects out the constants:
$P\m F=F-\int_{_{\NT^d}}F(\phi)\m d\phi$. 
Our strategy is to solve 
Eq.\,\,(\ref{fp1}) by an inductive RG method for given $\zeta$.
This turns out to be possible quite generally without any
nondegeneracy assumptions on $U$. The latter enter only in
the solution of Eq.\,\,(\ref{fp12}).
Below, we shall treat $W_0$ given by Eq.\,\,(\ref{W1})
as a map on a space of $\NR^{2d}$-valued functions $X$ 
on $\NT^d$ with arbitrary averages\footnote{That the 
solution $X$ of Eq.\,\,(\ref{fp1}) has zero average follows from 
the form of the equation.}. The vector $\zeta$ will be treated 
as a parameter and we shall often suppress it in the notation
for $W_0$.
\vskip 0.4cm

For the inductive construction of the solution 
of Eq.\,\,(\ref{fp1}), we shall decompose
\qq
G_0\ =\ G_1+\Gamma_0\m,  
\label{G2}
\qqq
where $\Gamma_0$ will effectively involve only the Fourier 
components with $|\omega\cdot q|$ larger than $\CO(1)$ 
and $G_1$ the ones with $|\omega\cdot q|$ 
smaller than that (see Sect.\,\,4). In particular, we shall 
have $\Gamma_0=\Gamma_0P$. Upon writing $X=Y+\tilde Y$,
\,Eq.\,\,(\ref{fp1}) becomes 
\qq
Y+\tilde Y\ =\ (G_1+\Gamma_0)\m P\m W_0(Y+\tilde Y)\m.
\label{dc}
\qqq
Suppose that $\tilde Y=\tilde Y_0$ where $\tilde Y_0$ solves 
for fixed $Y$ the ``large denominator'' equation:
\qq
\tilde Y_0\ =\ \Gamma_0 W_0(Y+\tilde Y_0)\m.
\label{lde}
\qqq
Then Eq.\,\,(\ref{dc}) reduces to the relation
\qq
Y\ =\ G_1P\m W_1(Y)
\label{fp2}
\qqq
if we define $W_1(Y)=W_0(Y+\tilde Y_0)$.
We have thus reduced the orginal problem (\ref{fp1})
to the one from which the largest denominators were
eliminated, at the cost of solving the easy large 
denominator problem (\ref{lde}) and of replacing 
the map $W_0$ by $W_1$. 
\vskip 0.4cm

Note that, with these definitions, $\tilde Y_0=\Gamma_0W_1(Y)$
and thus  $W_1$ satisfies the fixed point
equation
\qq
W_1(Y)\ =\ W_0(Y+\m\Gamma_0W_1(Y))\m.
\label{W2}
\qqq
Conversely, this equation, which we shall solve for $W_1$ 
by the Banach Fixed Point Theorem in a suitable space,
implies that $\tilde Y_0=\Gamma_0 W_1(Y)$ satisfies
Eq.\,\,(\ref{lde}) and thus that
\qq
X\,=\, Y+\m\Gamma_0W_1(Y)\,\equiv\, F_1(Y)
\label{F2}
\qqq
is a solution of Eq.\,\,(\ref{fp1}) if and only if
$Y$ solves Eq.\,\,(\ref{fp2}).
\vskip 0.4cm

After $n-1$ inductive steps, the solution of Eq.\,\,(\ref{fp1})
will be given as
\qq
X\,=\, F_{n-1}(Y)\m,
\label{Fn}
\qqq
where $Y$ solves the equation
\qq
Y\,=\, G_{n-1}P\m W_{n-1}(Y)
\label{fpn}
\qqq
and $G_{n-1}$ contains only the denominators
$\vert\omega\cdot q\vert\leq\CO(\eta^n)$
where $\eta$ is a positive number smaller than $1$ 
fixed once for all. The next inductive 
step consists of decomposing $G_{n-1}=G_{n}+\Gamma_{n-1}$ 
where $\Gamma_{n-1}$ involves $|\omega\cdot q|$  
of order $\eta^{n}$ and $G_{n}$ the ones smaller 
than that. We define $W_{n}(Y)$ as the solution 
of the fixed point equation
\qq
W_{n}(Y)\,=\, W_{n-1}(Y+\m\Gamma_{n-1} W_{n}(Y))
\label{Wn+1}
\qqq
and set
\qq
F_{n}(Y)\,=\, F_{n-1}(Y+\m\Gamma_{n-1} W_{n}(Y))
\label{Fn+1}
\qqq
(which is consistent with relation (\ref{F2}) if we take
$F_0(Y) = Y$).  
Then replacing $Y$ in Eqs.\,\,(\ref{Fn}) and (\ref{fpn})
by $\, Y+\m\Gamma_{n-1} W_{n}(Y)\m$, \m we infer that
$X=F_{n}(Y)$ if $Y=G_{n}P\m W_{n}(Y)$ completing 
the next inductive step. Note also the cumulative formulas 
that follow easily by induction:
\qq
&&W_n(Y)\,=\, W_0(Y\m+\m\Gamma_{<n} W_n(Y))\,,\label{cum1}\\
&&F_n(Y)\,=\, Y\m+\m\Gamma_{<n} W_n(Y)\,,\label{cum2}
\qqq
where $\Gamma_{<n}\m=\m\sum\limits_{k=0}^{n-1}\Gamma_k$ contains
all the denominators larger than  $\CO(\eta^n)$\m. 
\vskip 0.4cm

We shall control the transformations (\ref{Wn+1}) and (\ref{Fn+1})
in suitable norms. The point of the inductive procedure is
that $PW_n(Y)$ becomes effectively linear in $Y$ for large $n$, 
see Remark 1 after Proposition 3 below, so that $Y=0$ is a 
better and better approximation to a solution of the equation 
$Y=G_nPW_n(Y)$. In fact, as follows from the cumulative relations 
(\ref{cum1}) and (\ref{cum2}), $X_n\equiv F_n(0)=\Gamma_{<n}W_n(0)$ 
solves the approximate problem:
\qq
X_n\,=\,\Gamma_{<n}\m W_0(X_n)\m,
\label{apeq}
\qqq
obtained from Eq.\,\,(\ref{fp1}) by replacing $G_0$ 
by $\Gamma_{<n}$ (since $\Gamma_{<n}P=\Gamma_{<n}$).
We shall  construct the solution $X$ of Eq.\,\,(\ref{fp1}) as
the limit of the approximate solutions
\qq
X\,=\,\lim_{n\rightarrow\infty}\m X_n\m.
\label{theso}
\qqq

\nsection{Spaces}
\vs{2mm}

Let us rewrite the definition (\ref{W1}) in
terms of the Fourier transforms:  
$$
w_0(q;\m y)\,=\,\lambda \int_{\NT^d} \ee^{i\m q\cdot\phi}\,
\partial U((\phi,\zeta)+Y(\phi))\, d\phi\m,
$$
where we recall that $y$ refers to the Fourier transform of $Y$.
Let us explain here how to view $w_0$ as an analytic
functional on a suitable Banach space.
The analyticity  of $U$ implies the following. 
There exist $\rho>0$, $\alpha>0$ and $b<\infty$ such that 
the coefficients $U_{m+1}(\phi,\zeta)$, belonging to the space 
of $m$-linear maps $\CL(\NC^{2d},\dots,\NC^{2d};\NC^{2d})$, 
of the Taylor expansion  
$$
\partial U((\phi,\zeta)+Y)=\sum_{m=0}^\infty{_1\over^{m!}}\m 
U_{m+1}(\phi,\zeta)
(Y,\dots,Y)
$$
are analytic in $|\zeta|<\rho$ 
and their Fourier transforms satisfy the bounds 
\qq
\sum_q \ee^{\alpha |q|}\, \Vert u_{m+1}(q,\zeta)
\Vert_{_{\CL(\NC^{2d},
\dots,\NC^{2d};\NC^{2d})}}\leq\ b
\, m!\,\rho^{-m}\m.
\label{vmn}
\qqq
For later convenience, we shall use in $\NC^{2d}\cong
\NC^d\times\NC^d$ the norm $\vert\cdot\vert_{_0}$ defined by
\qq
\vert(z_1,z_2)\vert_{_0}\,\equiv\ \vert z_1\vert
+\vert z_2\vert
\label{deff}
\qqq
and the induced norms on the spaces of linear maps.
Inserting the Fourier series for $Y$ we end up
with the expansion
\qq
w_0(q\m ;\m y)&=&\sum\limits_{m=0}^{\infty} 
\sum_{\bf q}{_1\over^{m!}}\, u_{m+1}(q-\sum q_i,\m\zeta)
\m(y(q_1),\dots,y(q_m))\nonumber\\
&\equiv& \sum\limits_{m=0}^{\infty}
\sum_{\bf q}w_0^{(m)}(q,q_1,\dots,q_m;\zeta)\m(y(q_1),
\dots,y(q_m))\m,
\label{taylor}
\qqq
where ${\bf q}=(q_1,\dots,q_m)\in \NZ^{md}$.
This formula suggests to consider 
$w_0$ as an analytic function of $y$, where $y$ belongs 
to a suitable Banach space $h$. We take
$$h\m=\m\{\,y=(y(q))\ \vert\ y(q)\in\NC^{2d}\m,\ \,
{\Vert}y{\Vert}\equiv\sum\limits_{q}|y(q)|_{_0}\m
<\m\infty\,\}\m.$$ 
Let $B(r_0)$ be the open ball of radius $r_0$ in $h$ 
centered at the origin and let $H^\infty(B(r_0),h)$ denote
the Banach space of analytic functions \cite{cha}
$w:B(r_0)\rightarrow h$, equipped with the supremum 
norm, which we shall denote by $|||w|||$. 
The bound (\ref{vmn}) implies that $w_0\in H^\infty(B(r_0),h)$
for $r_0$ small enough, but before stating this, 
it is convenient to encode the decay property 
of the kernels $w_0^{(m)}$ inherited
from the estimate (\ref{vmn}) as a property 
of the functional $w_0$.
\vskip 0.4cm

For that let $\tau_\beta$  denote the translation 
by $\beta\in\NR^d$, $(\tau_\beta Y)(\phi)=Y(\phi-\beta)$. 
On $h$, $\tau_\beta$ is realized by $(\tau_\beta y)(q)
=e^{i\m \beta\cdot q}\m y(q)$. It induces a map $w\mapsto 
w_\beta$ from $H^\infty(B(r_0),h)$ to itself if we set  
$$
w_\beta(y)=\tau_\beta(w(\tau_{-\beta}y)).
$$ 
On the kernels $w^{(m)}$, this is given by 
\qq
w_\beta^{(m)}(q\m ;\m q_1,\dots,q_m)\,=\,\ee^{i\m
\beta\cdot(q-\sum q_j)}\, w^{(m)}(q\m ;\m q_1,\dots,q_m)\m.
\label{wbeta}
\qqq
and makes sense also for $\beta\in\NC^d$. We have 
$$
|||w_{0\beta}|||\,\leq\, \sum\limits_{m=0}^{\infty}
\sup_{q_1,\dots,q_m}\sum_{ q} \ee^{-{\rm Im}\m\beta 
\cdot (q-\sum q_j)}\m  
|w_0^{(m)}(q\m ;\m q_1,\dots,q_m;\zeta)|\, r_0^m
$$
Combining this with the bound (\ref{vmn}) we can summarize
the above discussion by

\vskip 0.6cm

\no{\bf Proposition 1}. {\it There exists $r_0>0$, 
$\alpha>0$ and  $D<\infty$,  such that 
$w_{0\beta}\in H^\infty(B(r_0),h)$ and it 
extends to an analytic function of $\beta$ in the region 
$|{\rm Im}\m\beta|<\alpha$ with values in $H^\infty(B(r_0),h)$ 
satisfying the bound
\qq
|||w_{0\beta}|||\,\leq\, D\m 
\vert\lambda\vert\m.
\label{wbeta1}
\qqq
Moreover, $w_{0\beta}$ is analytic in $\zeta$ for $|\zeta|<r_0$.}
\vs{6mm}

\no{\bf Remarks.} 1. \ The analyticity of $w_{0\beta}$ in a
strip in $\NC^d$ centered on $\NR^d$ followed from the exponential
decay of the kernels $w_0^{(m)}(q\m ;\m q_1,\dots,q_m)$ 
as functions of $q-\sum q_j$. In order to show that 
the solution of Eq.\,\,(\ref{fp1}) is real analytic in $\phi$,
we shall need to inductively establish such a decay for the kernels 
$w_n^{(m)}$ of the Fourier transforms $w_n$ of effective 
functionals $W_n$. This will follow once we establish 
the analyticity of $w_{n\beta}$ in $\beta$ with uniform bounds.
\vs{2mm}

\no 2. \ From now on, analyticity in $\zeta$ will always 
be understood to hold for $|\zeta|<r_0$.
\vs{4mm}

We finish this section by collecting, for convenience,
some standard properties of bounded analytic functions 
defined on open balls in Banach spaces (that are identical 
to those of analytic functions on finite dimensional
spaces, see \cite{cha}). Let $h\m,\ h'\m,\ h''$ be Banach spaces,
$B(r)\subset h$, $B(r')\subset h'$ and 
$w_i\in H^\infty(B(r),h')$, $w\in H^\infty(B(r'),h'')$. Then

\vskip 0.4cm

\no{\bf Composition property}. 
\no If $|||w_i
|||< r' $
then $w\circ w_i\in H^\infty(B(r),h'')$ and
\qq
|||w\circ w_i|||\,\leq\, |||w|||\m.
\label{circ}
\qqq
\vskip 0.4cm

\no{\bf Inequalities}. First of all, one deduces from the Cauchy 
estimate that for $r_1<r'$,
\qq
\sup_{\Vert x\Vert< r_1 } \Vert Dw(x)\Vert_{_{\CL(h';h'')}}
\leq (r'-r_1)^{-1} |||w|||\m,
\label{circ0}
\qqq
where ${\CL(h';h'')}$ denotes the space of bounded linear operators  
from $h'$ to $h''$. Taking $r_1=\hf r'$, we infer that if $|||w_i|||
\leq \hf r'$ then
\qq
|||w\circ w_1-w\circ w_2|||\,\leq\,\m {\frac{_2}{^{r'}}}
\,|||w|||\;|||w_1-w_2|||\m.
\label{circ1}
\qqq
Moreover,
if $\,\delta_k w(x) = w(x)
-\sum\limits_{\ell=0}^{k-1} \frac{1}{\ell !}\, D^\ell w(0)(x)$,
then
\qq
\sup_{\Vert x\Vert \leq \gamma\m r'}{\Vert}\delta_k 
w(x){\Vert}\,\leq\,\m 
{\frac{_{\gamma^k}}{^{1-\gamma}}} 
\m |||w|||
\label{circ2}
\qqq
for $0\leq \gamma < 1$.
\vskip 0.2cm

\nsection{Inductive bounds}
\vskip 0.2cm

In this section, we first define the operators $\Gamma_n$
used in  the RG transformations. Then, we state an easy result,
namely that the fixed point equations (\ref{Wn+1}) 
and (\ref{cum1}) may be solved for any $n$, if we choose
$\lambda$ small enough in an $n$-dependent way.
Then, we define the spaces in which our RG equations 
(\ref{Wn+1}) are eventually solved inductively for 
$\vert\lambda\vert$ small uniformly in $n$ and state precisely 
our inductive assumptions.
\vskip 0.4cm

In the Fourier variables, the fixed point equations (\ref{Wn+1}) 
and (\ref{cum1}) may  be written in the form 
\qq
&&\hbox to 1.3cm{$w_{n\beta}(y)$\hfill}
=\ w_{(n-1)\beta}(y\m+\m\Gamma_{n-1} 
w_{n\beta}(y))\m,
\label{wn+1}\\
&&\hbox to 1.3cm{$w_{n\beta}(y)$\hfill}
=\ w_{0\m\beta}(y\m+\m\Gamma_{<n}\m 
w_{n\beta}(y))\m,
\label{wnc}
\qqq
where we have introduced $\beta$ assuming that the operators 
$\Gamma_n$ are diagonal in Fourier space and hence
commute with $\tau_\beta$. Similarly, the equations 
(\ref{Fn+1}) and (\ref{cum2}) translate in the Fourier
space to the relations
\qq
&&\hbox to 1.3cm{$f_{n\beta}(y)$\hfill}
=\ f_{(n-1)\beta}(y\m+\m\Gamma_{n-1}\m 
w_{n\beta}(y))\m,
\label{fn+1}\\
&&\hbox to 1.3cm{$f_{n\beta}(y)$\hfill}
=\ y\m+\m\Gamma_{<n}\m 
w_{n\beta}(y)\m.
\label{fnc}
\qqq
\vskip 0.4cm

To define the operators $\Gamma_n$, we shall use an analytic 
partition of unity $(\chi_n)$ dividing the positive line 
into scales. Let  
\qq
\chi_0(\kappa)=1-\ee^{-\kappa^6}\m,
\quad\ \ \chi_n(\kappa)=\ee^{-(\eta^{-n+1}\kappa)^6}
-\,\ee^{-(\eta^{-n}\kappa)^6}
\ \ \quad{\rm for\ \ }n\geq1\m. 
\label{chi0}
\qqq
Clearly,
\qq
\hbox to 3.5cm{$\sum\limits_{n=0}^\infty\chi_n(\kappa)=1\m,$\hfill}
\chi_n(\kappa)=\chi_1(\eta^{-n+1}\kappa)
\quad{\rm for}\ \ n\geq 1\m.
\label{chi}
\qqq
Note that $\kappa^{-6}\m\chi_n(\kappa)$ are entire functions 
of $\kappa$ and that, for $|{\rm Im}\m\kappa|<B$ 
and $\ell=0,\dots,6$,
\qq
&&\vert\kappa^{-\ell}\m\chi_0(\kappa)\vert\,\leq\, C\m,
\label{chib0}\\
&&\vert\kappa^{-\ell}\m\chi_1(\kappa)\vert\,\leq\, C\,
\ee^{-\hf\m\vert\kappa\vert^6},
\label{chib1}
\qqq
for some $B$-dependent constant $C$.
Define 
\qq
\Gamma_n(q\m ,\m q')\,=\,\chi_n(\omega\cdot q)\,\m 
G_0(q\m ,\m q' )\,=\,
\gamma_n(\omega\cdot q)\,\m\delta_{q,q'}\m ,
\label{Gamma}
\qqq
where the matrix $\gamma_n$ is of the block form:
\qq
\gamma_n(\kappa)\,=\,\chi_n(\kappa)
\m\kappa^{-2}
\left(\matrix{\mu & 
i\kappa \cr 
-i\kappa & 0}\right)
\label{gamma}
\qqq
and we denote the kernel of an operator $a$ in $\CL (h;h)$ 
by $a(q,q')\in{\rm End}(\NC^{2d})$.
 We shall also need below more general operators 
$\Gamma_n(\kappa)$ with shifted kernels,
\qq
\Gamma_n(\kappa)(q,q')\,=\,\gamma_n(\omega\cdot q+\kappa)
\,\m\delta_{q,q'}\m.
\label{gnchi}
\qqq
It follows easily from the bounds (\ref{chib0}) and (\ref{chib1})
that for $n\geq 1$, 
\qq
\Vert\Gamma_{n-1}(\kappa)\Vert_{_{\CL(h;h)}}\m,
\ \Vert\Gamma_{<n}(\kappa)\Vert_{_{\CL(h;h)}}
\ \leq\ \ C\m\eta^{-2n}
\quad\quad{\rm if}\quad\ \ \vert\kappa\vert<\eta^{n} B
\label{<n0}
\qqq
with a new constant $C$ (say, twice bigger).
\vskip 0.4cm

Our goal is to show that $w_n$ and $f_n$ exist as 
analytic functionals provided that $|\lambda|$ is taken small
in an $n$-independent way. For later purposes it will be useful 
to prove this first for $|\lambda|$ small in an
 $n$-dependent way.
Although this is very easy to carry out,
it illustrates an important part of the main 
analytic argument in the general step.
\vskip 0.6cm

\no{\bf Proposition 2}. \ {\it For any sufficiently small
$r>0$, $\vert\lambda\vert<\lambda_{n}(r)$ and $|{\rm Im}\m\beta|
<\alpha$, the equations} (\ref{wnc}) {\it have a unique 
solution $w_{n\beta}\in H^\infty(B(r^n),h)$ with 
\qq
|||w_{n\m\beta}|||\ \leq\,D\vert\lambda\vert \m,
\label{w2<}
\qqq
where $D$ is as in Proposition 1.
The maps $f_{n\beta}$ defined by Eqs}.\,\,(\ref{fnc}) 
{\it belong to $H^\infty(B(r^n),h)$. They satisfy the bounds 
$|||f_{n\beta}|||\leq{2}\m r^n$. Moreover, $w_{n\beta}$ and 
$f_{n\beta}$ are analytic in $\lambda$, $\beta$ and $\zeta$ 
and they satisfy the recursive relations} (\ref{wn+1}) 
{\it and} (\ref{fn+1}){\it, respectively.}
\vskip 0.4cm

Postponing the proof to the end of the section,
we shall state the  bounds for $w_n$ that will be 
inductively  established for $\vert\lambda\vert$ small 
in an $n$-independent way. Due to the smallness of $\vert
\omega\cdot q\vert$ in the $n^{\rm th}$ scale, $\gamma_n$ will 
have very different 
effects in the variables $\theta$ and $j$ in $y=(\theta, j)$.
It will be therefore convenient to choose  $n$-dependent 
norms for $n\geq 1$. Let us first do it for $\NC^{2d}$
by defining
\qq
\vert(z_1,z_2)\vert_{_{\pm n}}\,\equiv\ \vert z_1\vert
+{_1\over^{\eta^{\pm n}}}\vert z_2\vert\m. 
\label{deff1}
\qqq
We shall use the notation $|\cdot|_{_{n;m}}$ for
the matrix norms induced by viewing a $2d\times 2d$ matrices
as maps from $\NC^{2d}$ with the norm $|\cdot|_{_n}$
to $\NC^{2d}$ with the norm $|\cdot|_{_m}$.
Next we set
\qq
\Vert y\Vert_{_n}
=\sum\limits_q|y(q)|_{_n}\,
\ee^{\eta^{-n}\vert\omega\cdot q\vert}\m.
\label{norm0}
\qqq
The weight $\ee^{\eta^{-n}\vert\omega\cdot q\vert}$ will
facilitate dealing with non-dangerous large denominators 
$\vert\omega\cdot q\vert$.
For $w$, it turns out to be useful to introduce the norms 
\qq
\Vert w\Vert_{_{-n}}
=\sum\limits_q|w(q)|_{_{-n}}\,
\ee^{-\eta^{-n}\vert\omega\cdot q\vert}\m.
\label{norm}
\qqq
Let $h_{\pm n}$ denote the corresponding Banach spaces. 
Note the natural embeddings for $n\geq 2$
\qq
h_n\ \longrightarrow\ h_{n-1}\ \longrightarrow\ h\,,
\quad\quad\quad\quad\quad h\ \longrightarrow
\ h_{-n+1}\ \longrightarrow\ h_{-n}\ \ 
\label{embedd}
\qqq
with the norms bounded by $1$:
\qq
\Vert\cdot\Vert\ \leq\ \Vert\cdot\Vert_{_{n-1}}\,
\leq\ \Vert\cdot\Vert_{_{n}}\,,
\quad\quad\quad\Vert\cdot\Vert_{_{-n}}\,\leq
\ \Vert\cdot\Vert_{_{-n+1}}\,\leq\ \Vert\cdot\Vert\,. 
\label{embe}
\qqq
For $n\geq2$ (but not for $n=1$), the operator 
$\Gamma_{n-1}$ or, more generally, operators 
$\Gamma_{n-1}(\kappa)$ may be considered as mapping
$h_{-n}$ into $h_n$. Indeed, it follows 
easily with the use of bound (\ref{chib1}) that 
\qq
\Vert\Gamma_{n-1}(\kappa)\Vert_{_{-n;n}}
\ \leq\ C\,\eta^{-2n}
\quad\quad{\rm if}\quad\ \ \vert\kappa\vert<\eta^{n-1}B
\label{5.5}
\qqq
with a new ($n$-independent) constant $C$.
\vskip 0.4cm

To simplify  notations, we shall denote by $B_n$ the open ball 
in  $h_n$ of radius $r^n$ 
and by $\CA_n$ the space $H^\infty(B_n,h_{-n})$
of analytic functions on $B_n$ with the supremum norm 
denoted by $|||\cdot|||$. 
Finally, for a linear operator $M:h_n\rightarrow h_{m}$ we use
the abbreviated notation $\Vert\cdot\Vert_{_{n;m}}$ for the norm 
in $\CL(h_{n},h_{m})$.
Due to the embeddings (\ref{embedd}), we may regard the maps 
$w_{n\beta}$, whose existence for sufficiently small 
$\vert\lambda\vert$ is claimed in Proposition 2, 
as belonging to $\CA_n$. 
Note that  both sides of  relation (\ref{wn+1}) are well
defined  for such maps due to the bound (\ref{5.5}) and
that their equality is implied by the results of Proposition 2. 
The next proposition states that, viewed as $\CA_n$-valued 
functions of $\lambda$, $w_{n\beta}$'s 
may be analytically extended to an $n$-independent 
disc $\vert\lambda\vert<\lambda_0$ (provided we restrict somewhat
the strip of $\beta$). It also lists the properties of such 
extensions.
\vskip 0.6cm

\no{\bf Proposition 3}. (a) \ {\it There exist positive constants
$r$ and $\lambda_0$ with $r<\eta^4$ such that, 
for $\vert\lambda\vert<\lambda_0$ and 
$\;\vert{\rm Im}\m\beta\vert< {\alpha_n}$, where} 
\qq
\alpha_1=\alpha\m,\quad \alpha_{n}=(1-n^{-2})\alpha_{n-1}\m, 
\quad n\geq2\m,
\label{an}
\qqq
{\it there exist solutions $w_{n\beta}\equiv w$ 
of Eqs.}\,\,(\ref{wn+1}) {\it belonging to $\CA_n$, analytic 
in $\lambda$, $\beta$ and $\zeta$
and coinciding with the solutions $w_{n\beta}$ of Proposition 2
for $\vert\lambda\vert<\lambda_n(r)$}.
\vs{2mm}

\no (b) \ {\it Writing 
\qq
w(y)=w(0)+Dw(0)y+\delta_2w(y),
\label{defs}
\qqq 
we have
\qq
&&{\Vert}P w(0){\Vert}_{_{-n}}\ \leq\ \epsilon\, r^{2n}\m,
\label{w0}\\
&&|||\delta_2w|||\quad\,\m \ \ \leq\ \epsilon\, r^{{3\over^2}n}\m,
\label{omega}
\qqq
where $\epsilon\rightarrow 0$ as $\lambda\rightarrow 0$}.
\vs{2mm}

\no(c) \
\qq
\quad{\Vert}Dw(y){\Vert}_{_{n;-n}}\ 
\leq\ \epsilon\,\eta^{2n}\m.
\label{Dwbound}
\qqq

\vskip 0.5cm

\no{\bf Remarks.} 1. \ If we rescale the maps $w_n$ by
introducing $\m\tilde w_n(y)\,=\, \eta^{-2n}\m r^{-n}\, w_n(r^ny)\m$
then it follows from the above statements that 
$\tilde w_{n\beta}\equiv\tilde w$ are analytic maps from
a unit ball in $h_n$ to $h_{-n}$ and $\tilde w(y)
=\tilde w(0)+ D\tilde w(0) y+\delta_2\tilde w(y)$ with
\qq
\Vert P\tilde w(0)\Vert_{_{-n}}\m\leq\m\epsilon\,\eta^{-2n}\m r^n\m,
\quad\ \vert\vert\vert\delta_2\tilde w\vert\vert\vert\m\leq
\m\epsilon\, \eta^{-2n}\m r^{\hf n}\m,\quad\ \Vert D\tilde w(0)
\Vert_{_{n,-n}}\m\leq\m\epsilon\m.
\nonumber
\qqq
Hence with growing $n$, $\m P\tilde w$ becomes an approximately 
linear map.
\vskip 0.3cm

\noindent 2. \ Let us explain the idea of the proof of Proposition 3. 
Consider the linearization of Eq.\,\,(\ref{wn+1}):
\qq 
w_{n}=w_{n-1}+Dw_{n-1}\Gamma_{n-1} w_n\, +\ \dots
\label{linr}
\qqq
In order to solve the above equation one has to invert 
the operator $1-Dw_{n-1}\Gamma_{n-1}$: 
$$\m w_n=(1-Dw_{n-1}\Gamma_{n-1})^{-1}\m w_{n-1}\,+\,\dots$$ 
However, operator $\Gamma_{n-1}$ is of order $\eta^{-2n}$  
as a map from $h_{-n}$ to $h_{n}$ (recall the bounds 
(\ref{5.5})) and we need to show that $Dw_{n-1}$ is effectively 
of order $\eta^{2n}$ as a map from $h_n$ to $h_{-n}$, which is, 
essentially, what Eq.\,\,(\ref{Dwbound}) says with $n$ shifted 
to $n-1$. Altogether, $Dw_{n-1}\Gamma_{n-1}$ remains of order 
$\epsilon$ as a map from $h_{-n}$ to $h_{-n}$ 
(this motivates also our choice of the norms)
and $\Vert(1-Dw_{n-1}\Gamma_{n-1})\Vert_{_{n;-n}}\leq
1+\CO(\epsilon)$. In the proof of the estimate (\ref{Dwbound}),
we shall need the Ward identities discussed in Sect.\,\,5. 
This is the only subtle part of our argument.
Indeed, once the bound (\ref{Dwbound}) is shown, the rest 
of the proof of Proposition 3 reduces to the standard Banach 
Fixed Point Theorem combined with the Diophantine 
property of $\omega$.
\vskip 0.2cm

The latter is used in the following way (which is 
similar to the way it enters the standard
KAM proof): upon iteration, we consider smaller and smaller
$\vert\omega\cdot q\vert\m$'s, of order $\eta^n$. This means 
$|q|$ is of order $\eta^{n/\tau}$, by the Diophantine
condition (\ref{Dio}).  On the other hand, the introduction 
of the parameter $\beta$ in (\ref{wbeta}) allows to preserve 
the exponential decay of the kernels $w_0^{(m)}(q; q_1, 
\dots, q_m)$ in the size of $|q-\sum q_j|$. By shrinking 
at each step the analyticity region in $\beta$ we show that 
the leading contribution to $w_n$'s given by $w_n(0)$ 
(see Eq.\,\,(\ref{defs})) contracts for $q\neq 0$.
Actually, ${\Vert}Pw_n(0){\Vert}_{_{-n}}$ decays 
super-exponentially in $n$, see the estimate (\ref{omega.q}) below,
which explains why we can choose $r$ as small as we want.
\vskip 0.2cm

Finally, the bound (\ref{omega}) is easy to understand.
By definition, $\delta_2w_n$ and its first derivative vanish 
for $y=0$, and the norm $|||\delta_2w_n|||$ 
is defined by taking the supremum over balls of radius $r^n$, hence 
one expects $|||\delta_2w_n|||$ to be of order $(1+\CO(\epsilon))^n
\m r^{2n}$ by the  Cauchy estimate (\ref{circ2}) (the weaker bound 
(\ref{omega}) is sufficient and is a convenient way to control 
the non-linear corrections to the iteration). Recall that, 
eventually, we  construct our solution as a limit of $X_n =F_n(0)$, 
for which we need to control $w_n(y)$ only for $y=0$,
see Eq.\,\,(\ref{cum2}). Thus we can let the radius $r^n$ 
of the ball where our estimates hold tend to zero.

\vs{2mm}

\no 3. \ Combining all the bounds, we get 
\qq
|||w_n-(1-P)w_n(0)|||\ 
\leq\ C\,\epsilon\,\eta^{2n}\m \m.
\label{wbound}
\qqq
The zero mode part $(1-P)w_n(0)$ 
of $w_n(0)$ will be controlled later, see Eqs.\,\,(\ref{wi1})
and the second of Eqs.\,\,(\ref{smll}) below from which it follows
that it is of the form $(0,\xi_n)$ where $\xi_n=\CO(\lambda)$ 
converges in $\NR^d$ when $n\to\infty$. Note that, since $w_n$
is multiplied by $\Gamma_{n-1}= \Gamma_{n-1} P$ in the argument 
of $w_{n-1}$ in Eq.\,\,(\ref{wn+1}), the constant mode $(1-P)w_n(0)$ 
may be decoupled from the iteration and we do not need to control 
it in order to prove Proposition 3.
\vs{2mm}

\no 4. \ We choose the constants as follows.
$\eta<1$ has been fixed first. $B$ which enters the estimates
(\ref{<n0}) and (\ref{5.5}) is chosen then large enough (see 
Eq.\,\,(\ref{jb2}) below).
Given those, $r$ and then $\lambda_0$ are chosen small enough. 
It should be emphasized that all quantities
that are bounded by an $n$-dependent power of $r$ are easy 
to estimate and that these estimates do not involve the Ward 
identities, the latter entering only in bounds with 
$\eta^{2n}$. Finally, we denote by $C$ a generic constant, 
independent of $\epsilon$ and $n$, which may vary from place 
to place.

\vs{6mm}

Let us end this section with the easy

\vskip 0.4cm
 
\no{\bf Proof of Proposition 2.} \ Consider the
fixed point equation (\ref{wnc})
and write it as $w=\CF(w)$ for $w=w_{n\beta}$ and
$$
\CF(w)(y)=w_{0\beta}(y+\Gamma_{<n} w(y)) .
$$
Let $\CB_n$ denote the closed ball composed of $w
\in H^\infty(B(r^n),h)$ with $|||w|||\leq  D\lambda_{n}$
(where $D$ is as in Proposition 1).
Choose $\lambda_{n}$ so that
$C \m\eta^{-2n}\m D\lambda_{n}\leq r^n$
with $C$ as in the bounds (\ref{<n0}).
It follows from the latter that for $w\in \CB_n$
and $y\in B(r^n)\subset h$, 
\qq
\m{\Vert}y+ \Gamma_ {<n}w(y){\Vert}\leq
r^n\m+\m
C\m\eta^{-2n}\m D\lambda_{n}\leq 2r^n
\leq \hf\m r_0
\label{jb}
\qqq
for $r$ sufficiently small. Thus $\CF(w)$ is defined in $B(r^n)$ 
and, by Proposition 1, $\Vert\CF w(y)\Vert\leq D\vert\lambda\vert
\leq D\lambda_n$. Hence $\CF:\CB_n\rightarrow\CB_n$. 
For $w_i\in \CB_n$, $i=1,2$, use the property 
(\ref{circ1}) to conclude that 
$$|||\CF(w_1)-\CF(w_2)|||\leq
{_2\over^{r_0}}\m C\m\eta^{-2n}\m D\lambda_n\m|||w_1-w_2|||\leq
{_{2\m r^{n}}\over^{r_0}}\m|||w_1-w_2|||\leq\hf\m|||w_1-w_2|||
$$ 
for $r$ as in the estimate (\ref{jb}), i.e. that $\CF$ is 
a contraction. It follows that Eq.\,\,(\ref{cum1})
possesses a unique solution $w_{n\beta}$ in $\CB_n$
satisfying the bound (\ref{w2<}) which, 
besides, is analytic in $\lambda$, $\beta$ and $\zeta$.
Consider now for $n\geq 2$ 
the map $\CF'$:
\qq
\CF'(w)(y)= w_{0\m\beta}(\m y\m+\m\Gamma_{n-1}\m 
w_{n\beta}(y)\m+\m\Gamma_{<n-1} w(y))\m.
\label{n1}
\qqq
Again, $\CF'$ is a contraction in $\CB_{n}$
since, for $\Vert y\Vert\leq  r^{n}$, 
we have $\Vert y+\m\Gamma_{n-1}\m 
w_{n\beta}(y)\m+\m\Gamma_{<n-1} w(y)\Vert\leq
3r^{n}\leq \ha r_0$ for $r$ sufficiently small.
But Eqs.\,\,(\ref{wnc}) imply that 
$w_{n\beta}$ and $w_{(n-1)\beta}\circ(\m 1+
\Gamma_{n-1} w_{n\beta})$, both in $\CB_{n}$, are its fixed 
points and, consequently, they have to coincide. Hence 
the recursions (\ref{wn+1}) follow. By virtue of the estimate
(\ref{jb}), $\m{\Vert}y+ \Gamma_ {<n}w(y){\Vert}\leq 2r^n$
for $y\in B(r^n)$. By definition (\ref{fnc}), this gives 
the claimed bound on $|||f_{n\beta}|||$. The recursion 
(\ref{fn+1}) follows easily from Eq.\,\,(\ref{wn+1}).
\ \ $\Box$

\nsection{Ward identities and cancellation of resonances}
\vskip 0.2cm

The goal of this section is to prove the properties of the maps 
$W_n$ which will be essential in the proof of part (c)
of  Proposition 3, see Lemma 2 below, and in the proof 
Theorem 1, see Lemma 1. These properties, which are proven 
by a simple integration by parts, result from the symmetries 
of $W_0$ and will be encoded in the identities which, in the QFT 
formulation of the problem explained in Sect.\,\,8, can be 
interpreted as the {\it Ward identities} corresponding 
to the translation symmetry. Their function in the proof is 
to guarantee a partial cancellation of the repeated resonances 
that plague the Lindstedt series, see Sect.\,\,9.
\vskip 0.3cm

Indeed, as we shall see, the subtle part of the proof 
of the estimate (\ref{Dwbound}) reduces to a bound 
on the diagonal elements of the kernel $Dw_n(0)(q,q)$ of 
the derivative $Dw_n$ evaluated at zero. Our strategy will 
be to show that this kernel is actually a function
of $\omega\cdot q$ only and is of the form 
$\,(\matrix{_{\CO((\omega\cdot q)^2)} & _{\CO(\omega\cdot q)}\cr 
^{\CO(\omega\cdot q)} & ^{\CO(1)}})\,$ for $\omega\cdot q$
small. This, combined with our choice of the norms (\ref{deff1})
and (\ref{norm}), will then be used to imply the estimate
(\ref{Dwbound}). To show such a behavior of  $Dw_n(0)(q,q)$,
we shall compute certain of its  derivatives 
at $\omega\cdot q=0$  and show that they vanish.
This is the role of the Ward identities proven here.
Since $q$ is in ${\bf Z}^d$, to make sense of such derivatives
we need to introduce a smooth interpolation of
$Dw_n(q,q)$ viewed as a function of $\omega\cdot q$. 
This is the role of the functions $\pi_n$ defined after Lemma 1 
below.
\vskip 0.4cm

For simplicity, we shall first state and prove the Ward identities 
for the maps $W_n$ constructed in Proposition 2 for 
$\vert\lambda\vert<\lambda_n$ (by analyticity in $\lambda$,
they will also hold for $W_n$'s which will be constructed 
in Proposition 3 for $\vert\lambda\vert<\lambda_0$). 
The basic identity reads (recall that $Y^i=\Theta^i$ for $i\leq d$)
\qq
\int_{\NT^d} W_n^i(\phi;Y)\, d\phi\,=\,
\int_{\NT^d} Y^l(\phi)\,\,\partial_{\phi^i} 
W_n^l(\phi;Y)\,\, d\phi \;\quad{\rm for}\ \ i\leq d
\label{ward}
\qqq
or, in the Fourier language,
\qq
w_n^i(0\m ;\m y)\, =\, i\sum\limits_{q} 
y^l(q\m )\, q^i\, w_n^l(-q\m ;\m y)\m, 
\label{wardf}
\qqq
where on the right hand sides the summations over the repeated
index $l$ from $1$ to $2d$ are understood.
Let us check first the $n=0$ case, see Eq.\,\,(\ref{W1}),
\qq
\int(\partial_i U)((\phi,\zeta)+Y(\phi))\,\m d\phi&=&
\int\partial_i \m[\m U((\phi,\zeta)+Y(\phi))\m]\,\, d\phi\cr
&-&\int(\partial_l U)((\phi,\zeta)+Y(\phi))\,\m
\partial_iY^l(\phi)\,\, d\phi\m.
\nonumber
\qqq                             
The first term on the right hand side vanishes and 
the second one yields the claim by integration by parts. 
For $n\geq1$, using the relation (\ref{cum1}), we obtain
\qq
\int W^i_{n}(\phi;Y)\,\m d\phi 
&=& \int W_0^i(\phi;Y+\Gamma_{<n}W_{n}(Y))\,\m d\phi\cr
&=&\int(\m Y^l+(\Gamma_{<n} W_n)^l(Y)\m)(\phi)\,\,
\partial_{\phi^i}W_0^l(\phi; Y+\Gamma_{<n} W_{n}(Y))\,\m d\phi\cr
&=&\int Y^l(\phi)\,\partial_{\phi^i} 
W^l_{n}(\phi;Y)\, d\phi\,+\m\int(\Gamma_{<n} 
W_{n}(Y))^l(\phi)\,\m\partial_{\phi^i}W^l_{n}(\phi;Y)\, 
d\phi\m.
\nonumber
\qqq
The last integral can be written as
$\int\Gamma_{<n}^{l\m{l}'} (\phi - \phi')\, 
W^{{l}'}_{n}(\phi',Y)\,
\partial_{\phi^i}W^l_{n}(\phi;Y)\, d\phi\,d\phi'$ 
and two integrations by parts and the symmetry of $\Gamma_{<n}$
show that it is equal to its opposite, hence that it vanishes.
\vskip 0.4cm

To derive the consequences
of the  Ward identity (\ref{wardf}) used later,
 evaluate  it first at $y=0$: 

\vskip 0.7cm

\no{\bf Lemma 1}. 
\vskip -1.25cm
\qq
w_{n}^i(0;\m y)\vert_{y=0}\,=\ 0\qquad{\rm for\ all}
\quad n\quad {\rm and} \quad i\leq d\m. \label{wi1}
\qqq
\vskip 0.4cm

The next identities involve $Dw_n$. Let
us first introduce smooth interpolations of 
the diagonal parts of the kernels of the derivatives 
$Dw_{n\beta}$ of the maps $w_{n\beta}$ constructed in 
Proposition 2 for $\vert\lambda\vert<\lambda_{n}$.
These derivatives are given by the formula
\qq
Dw_{n\beta}(y)\m=\m[\m1 -\m Dw_{0\beta}(y_n)\,
\Gamma_{<n}\m]^{-1}\m Dw_{0\beta}(y_n)
\label{dww}
\qqq
with $y_n\equiv y+\Gamma_{<n} w_{n\beta}(y)$.
This is obtained from the $y$-derivative of Eq.\,\,(\ref{wnc}).
We shall now proceed to show that
the kernel $Dw_{n\beta}(q,q';\m y)$ on the diagonal $q=q'$
depends on $q$ only through $\omega\cdot q$.
For this purpose, let us introduce the continuous automorphism  
$t_p:\CL(h;h)\rightarrow\CL(h;h)$, for $p\in \NZ^d$,
shifting both arguments of the kernel of an operator $a$ by $p$:
\qq
(t_p\m a)(q,q')\m=\m a(q+p,q'+p)\m.
\label{auto}
\qqq
For $n=0$, $t_p\m Dw_0=Dw_{0\beta}$, as follows from the 
explicit form (\ref{taylor}) of the Taylor coefficients of $w_0$.
Applying $t_p$ to Eq.\,\,(\ref{dww}), we obtain the relation
\qq
t_p\m Dw_{n\beta}(y)\m=\m[\m1 -\m Dw_{0\beta}
(y_n)\,
t_p\m\Gamma_{<n}\m]^{-1}\, Dw_{0\beta}(y_n)\m.
\label{dwwp}
\qqq
Note that $t_p\Gamma_{<n}=\Gamma_{<n}(\omega\cdot p)$, 
see the definition (\ref{gnchi}). 
It follows that $t_p\m Dw_{n\beta}$ depends on $p$ only through 
the scalar product $\omega\cdot p$. Denote $Dw_{0\beta}(y)
=\pi_{0\beta}(y)$ and 
define for $n\geq1$ and $\vert\kappa\vert<\eta^{n}\m B$,
\qq
\pi_{n\beta}(\kappa;\m y)\ =\ [\m1-\,\pi_{0\beta}(y_n)\,\Gamma_{<n}
(\kappa)\m]^{-1}\m\pi_{0\beta}(y_n)\m. 
\label{pinn}
\qqq
Since, by the inequalities (\ref{jb}), 
$\Vert y_n\Vert\leq\ha r_0$ for
$y\in B(r^n)\subset h$, Proposition 1 and the Cauchy 
estimate (\ref{circ0}) imply that $\Vert\pi_{0\beta}(y_n)
\Vert_{_{\CL(h;h)}}\leq\,{2\over r_0}\m D\m\vert\lambda\vert$.
It follows then easily that $\pi_{n\beta}(\kappa;\m y)$ is analytic
for $\vert\lambda\vert<\lambda_n(r)$, 
$\vert{\rm Im}\m\beta\vert<\alpha$, $\vert\kappa\vert<\eta^{n} B$ 
and $y\in B(r^n)\subset h$ with the norm arbitrarily small 
if $\lambda\to 0$, e.g. 
\qq
\Vert\m\pi_{n\beta}(\kappa;\m y)
\Vert_{_{\CL(h;h)}}\leq
\ \vert\lambda\vert^{1/2}\m.
\label{pio}
\qqq
Comparing Eqs.\,\,(\ref{dwwp}) and (\ref{pinn}), we infer that
\qq
t_p\m Dw_{n\beta}(y)=\pi_{n\beta}
(\omega\cdot p;\m y)\quad\ {\rm and}\quad\ 
t_p\pi_{n\beta}(\kappa)=\pi_{n\beta}(\kappa+\omega\cdot p)\m,
\label{pioo}
\qqq
whenever defined, i.e. that $\pi_{n\beta}(\kappa)$ is a smooth 
interpolation of $t_p\m Dw_{n\beta}$. 
Note an explicit expression, which we shall need later, 
for the $\kappa$-derivative of $\pi_{n\beta}(\kappa)$ obtained 
by differentiating Eq.\,\,(\ref{pinn}):
\qq
\da_\kappa\pi_{n\beta}(\kappa;\m y)\,=\,\pi_{n\beta}(\kappa;\m y)
\,\da_\kappa\Gamma_{<n}(\kappa)\,\pi_{n\beta}(\kappa;\m y)\m.
\label{1dpi}
\qqq
As is easy to check, the maps $\pi_{n}$ satisfy for 
$\Vert y\Vert< r^n$ and $\vert\kappa\vert<\eta^{n}B$ 
the recursion relation
\qq
\pi_{n\beta}(\kappa;\m y)\ =\ [\m1-\,\pi_{(n-1)\beta}
(\kappa;\m{\tilde y})\,\Gamma_{n-1}(\kappa)\m]^{-1}
\,\pi_{(n-1)\beta}(\kappa;\m {\tilde y})\m,
\label{rinn}                                            
\qqq
where we have denoted $\tilde y\equiv y+\Gamma_{n-1}w_{n\beta}(y)$.
\vskip 0.5cm

The second consequence of the Ward identity is
\vskip 0.4cm

\no{\bf Lemma 2}.
\vskip -1.25cm
\qq
\pi_{n}^{ij}(0,0;\m\kappa;\m y)\vert_{\kappa=0\atop y=0}\ \ \ 
&=&0\;\quad{\rm for}\ \ i\;{\rm or}\;j\m\leq\m d\label{l=00}\\
\partial^\ell_\kappa\pi_{n}^{ij}(0,0;\m\kappa;\m y)
\vert_{\kappa=0\atop y=0}&=&0 
\;\quad{\rm for}\ \ i\;{\rm and}\;j\m\leq d\quad{\rm and}\ 
\ell\m<\m2\m.
\label{l=01}
\qqq

\vs{1mm}

\no {\bf Remark.} Eqs.\,\,(\ref{l=00}), (\ref{l=01})
and  (\ref{pio}) 
 imply that 
$\pi_n(0,0;\m\kappa;\m y) \vert_{y=0}=\,
(\matrix{_{\CO(\kappa^2)} & _{\CO(\kappa)}\cr 
^{\CO(\kappa)} & ^{\CO(1)}})$, the fact which will be used 
in an essential way in the next section.
\vskip 0.4cm

\no {\bf Proof.} First, evaluating the derivative 
of Eq.\,\,(\ref{wardf}) w.r.t. $y$ at $y=0$, we obtain:
\qq
Dw_n^{il}(0,\m q\m ;\m 0)\,=\,i\m {q}^{\m i}\, w_n^l(-q\m ;\m 0)
\quad\quad{\rm for}\ \ i\leq d\m.
\label{2point}
\qqq
\vs{4mm}
The kernels of $Dw_n$ possess the symmetry property
\qq
Dw_n^{l\m{l}'}(q\m ,\m q';\m y)\,=\, Dw_n^{{l}'l}
(-q',-q\m ;\m y)
\label{sz}
\qqq
which follows by Eq.\,\,(\ref{dww}) from the similar 
property of $Dw_0^{l\m{l}'}(q,q';\m y)$.
The latter comes from  the fact 
that $Dw_0$ is the symmetric second derivative 
of the functional $\lambda\int U((\phi,\zeta)+Y(\phi))\m d\phi$ 
and Eq.\,\,(\ref{sz}) encodes, at least formally, the analogous 
property of $W_n$, see Sect.\,\,8. More generally,
\qq
\pi_{n}^{l\m{l}'}(q\m ,\m q';\m\kappa;\m y)\,=\, \pi_{n}^{{l}'l}
(-q',-q\m ;-\kappa;\m y)\m,
\label{sz1}
\qqq
where we denote the kernels of the maps 
$\pi_n(\kappa;\m y)$ by $\pi_{n}(q,q';\m \kappa;\m y)$.
Setting $q=0$ in Eq.\,\,(\ref{2point}) and using 
the symmetry (\ref{sz}), we obtain the first claim
\qq
Dw_n^{ij}(0,0;\m y)\vert_{y=0}\,=\ 
\pi_{n}^{ij}(0,0;\m\kappa;\m y)\vert_{\kappa=0\atop y=0}
\,=\ 0\quad\;{\rm for}\ \ i\ {\rm or}\ j\m
\leq \m d\m.
\label{dw000}
\qqq
For the second claim, we use the relation
(\ref{1dpi}) for $\da_\kappa\m\pi_n$ which, written
in terms of the kernels $\pi_n(q,q';\m\kappa;\m y)$, yields:
\qq
\da_\kappa\pi_n(q,\m q';\m\kappa;\m y)\vert_{\kappa=0\atop y=0}\,
=\ \sum\limits_{q''} Dw_n(q,\m q''\m;\m 0)
\m\,\da_\kappa\gamma_{<n}(\omega\cdot q'')\, 
Dw_n(q''\m,\m q';\m 0)\m.
\nonumber
\qqq
In particular, for $q=q'=0$,
\qq
\da_\kappa\pi_n^{ij}(0\m ,\m 0\m ;\m\kappa;\m y)
\vert_{\kappa=0\atop y=0}
\,=\ \sum\limits_{q} Dw_n^{il}(0\m ,\m q;\m 0)
\,\m\da_\kappa\gamma_{<n}^{l\m{l}'}(\omega\cdot q)\, 
Dw_n^{j{l}'}(0\m ,-q\m;\m 0)\m,
\nonumber
\qqq
where we have also used the symmetry (\ref{sz}).
Finally, for $i,j\leq d$, substituting the  Ward identity 
(\ref{2point}), we obtain the relation
\qq
\da_\kappa\pi_n^{ij}(0,\m 0;\m\kappa;\m y)\vert_{\kappa=0\atop y=0}
\,=\ \sum\limits_{q}q^i\m q^j\,\m  w_n^l(-q;\m 0)\, 
\da_\kappa\gamma_{<n}(\omega\cdot q)^{l\m{l}'}
w_n^{{l}'}(q;\m 0)\m.
\label{dpi}
\qqq
The right hand side of the last expression is symmetric 
in indices $i,\m j$ but, by the  symmetry (\ref{sz1}), the left 
hand side is antisymmetric, hence zero. We can also see this 
more directly since, as follows from Eq.\,\,(\ref{gamma}),
$\da_\kappa\gamma_{<n}(\kappa)=(\matrix{_{a(\kappa)}&_{b(\kappa)}\cr 
^{-b(\kappa)} &^0})$ with $a$ odd and $b$ even. Thus
the expression summed on the right hand side of Eq.\,\,(\ref{dpi}) 
is odd in $q$ and the sum vanishes. 
\ \ $\Box$

\nsection{ Proof of Proposition 3}  
\vskip 0.2cm

\no The proof contains two different parts.
The inductive proofs of parts (a) and (b) are
straightforward applications of the Banach
Fixed Point Theorem and the contraction in $n$
follows easily by combining analyticity with  the Diophantine
condition. The proof of part (c) may also be divided into two 
parts. First, one controls $Dw_n(y)-Dw_n(0)$ and the off-diagonal
elements of $\m Dw_n(0)(q,q')$. This is also
straightforward and similar to the proofs in parts (a) and (b).
For $Dw_n(0)(q,q)$, we use Lemma 2 above. 
\vskip 0.4cm

For $n\leq n_0$ with fixed $n_0$, the bounds (\ref{w0})
and (\ref{omega}) as well as (\ref{Dwbound}) follow immediately 
from the estimate (\ref{w2<}) and the inequalities 
$\m\Vert\cdot\Vert\leq\Vert\cdot\Vert_{_n}\m$,\ 
$\m\Vert\cdot\Vert_{_{-n}}\leq\m\Vert\cdot\Vert\m$
by taking $\lambda$ small enough. For $n>n_0$ with $n_0$ large 
enough, we shall proceed inductively. It will be convenient 
to modify slightly the simplified notations of the text 
of Proposition 3 and so, below, $w$ will stand 
for $w_{(n-1)\beta}$ and $w'$ for $w_{n\beta}$. Finally, 
$\Gamma$ will stand for $\Gamma_{n-1}$.
\vskip 0.4cm

\no{\bf Proof of (a)}. \ Consider the recursive equation
(\ref{wn+1}) for $w'$ and use the decomposition (\ref{defs}) 
to rewrite it as
$$
w'(y)=w(0)+Dw(0)(y+\Gamma w'(y))+\delta_2w(y+\Gamma w'(y))
$$
from which we deduce that
\qq
w'(y)=Hw(0)+HDw(0)y+u(y)\m,
\label{w'}
\qqq
where
\qq
u(y)\equiv H\delta_2w(y+\Gamma w'(y))=H\delta_2w(\Gamma Hw(0)
+{\tilde H}y+\Gamma u(y))
\label{u}
\qqq
with  $H=(1-Dw(0)\Gamma)^{-1}$ and ${\tilde H}=1+\Gamma\m 
H\m Dw(0)=(1-\Gamma\m Dw(0))^{-1}$\m. 
\vskip 0.4cm

In the inductive step, first we assume 
that $w$ satisfies the bounds (\ref{w0}), (\ref{omega}) and 
(\ref{Dwbound}) with $n$ replaced by $n-1$. The bounds 
(\ref{Dwbound}), (\ref{5.5}) and (\ref{embe}) imply then 
that the operators $H$ and $\tilde H$ are well defined with
\qq
&{\Vert}H{\Vert}_{_{-n+1;-n+1}}\; ,\;{\Vert} {\tilde H} 
{\Vert}_{_{n-1;n-1}}
\ \leq\ 1+C\epsilon\ \leq\ 2,
\label{H2}
\qqq
for $\epsilon$ (i.e.\,\,$\vert\lambda\vert$) small enough.
\vskip 0.4cm

We solve Eq.\,\,(\ref{u}) using the Banach Fixed Point Theorem.
Given its solution $u$, the existence of $w'$ satisfying 
Eq.\,\,(\ref{w'}) follows. To solve Eq.\,\,(\ref{u}), we 
consider the map $\CG$ 
defined by
\qq
\CG(u)(y)=H\delta_2w(\tilde y)\quad\ \ {\rm with}\quad\ \ 
\tilde y=\Gamma Hw(0)+{\tilde H}y+\Gamma u(y)\m.
\label{2}
\qqq
We  claim that $\CG$ is a contraction in the ball
\qq
\CB=\{u\in H^\infty(B_\delta,h_{-n+1})\;|\m\;|||u|||\leq
2\m\epsilon\m r^{{_3\over^2}(n-1)}\}\m,
\label{ball}
\qqq
where ${B_\delta}\subset h_{n-1}$ is the open ball of
radius $r^{n-\delta}$ for $0\leq\delta<1$ and  $r<r(\delta)$. 
Indeed, for ${\Vert}y{\Vert}_{_{n-1}}\leq r^{n-\delta}$, the inequalities  
(\ref{5.5}), (\ref{H2}) and (\ref{w0}) combined with the
identity $\Gamma H w(0)=\Gamma H\m P w(0)$
($HP=H$ follows from $\Gamma \m P=\Gamma$ and the definition
of $H$)
and the inequalities (\ref{embe})
imply that for $u\in\CB$,
\qq
{\Vert}\tilde y{\Vert}_{_{n-1}}\leq  
2\m C\m\eta^{-2n}\m\epsilon\m r^{2(n-1)}
+2\m r^{n-\delta}+2\m C\m\eta^{-2n}\m\epsilon\m
r^{{_3\over^2}(n-1)}
\leq\, \hf\m r^{n-1}                            
\label{11}
\qqq
if $r$ is small enough. Thus $\tilde y\in B_{n-1}$, 
the domain of definition of $w$ and hence of $\delta_2w$. 
It follows that $\CG(u)$ for $u$ in the ball (\ref{ball}) 
may be considered as an analytic map of $y\in {B_\delta}$ with values 
in $h_{-n+1}$. Moreover
\qq
{\Vert}\CG(u)(y){\Vert}_{_{-n+1}}\,\leq\, 2\m |||\delta_2w||| 
\,\leq\, 2\m\epsilon\m r^{{_3\over^2}(n-1)}\m,
\label{3}
\qqq
where we used the bounds (\ref{omega}), (\ref{embe}) 
and (\ref{H2}). Hence $\CG :\CB\rightarrow\CB$. 
\vskip 0.4cm

To prove that $\CG$ is a contraction, we use the estimate 
(\ref{circ1}) for $\tilde y_i (y) =\Gamma Hw(0)+{\tilde H}y
+\Gamma u_i(y)$ and $u_i\in\CB$, $i=1,2$. By  inequality (\ref{11}), 
$\Vert\tilde y_i\Vert_{_{-n+1}}\leq\ha r^{n-1}$ 
and so the bounds (\ref{circ1}), (\ref{omega}), (\ref{embe}) 
and (\ref{H2}) imply that
\qq
&&|||\CG(u_1)-\CG(u_2)|||
= \sup\limits_{y\in {B_\delta}}\Vert H\delta_2w(\tilde y_1)
-H\delta_2w(\tilde y_2)\Vert_{_{-n+1}}\cr
&&\leq 4\m r^{-n+1}\m |||\delta_2w|||
\,\sup\limits_{y\in {B_\delta}}\Vert\tilde y_1-\tilde y_2
\Vert_{_{n-1}}\leq 4\m\epsilon\, r^{{1\over2}(n-1)}\,
\sup\limits_{y\in {B_\delta}}\Vert\tilde y_1-\tilde y_2
\Vert_{_{n-1}}\cr 
&&\leq 4\m\epsilon\, r^{{1\over 2}(n-1)}\m C\m\eta^{-2n}\, 
|||u_1-u_2|||\leq \hf |||u_1-u_2|||
\nonumber
\qqq
for $r$ and $\epsilon$ small proving the contractive property 
of $\CG$ on $\CB$. 
Hence the existence of the fixed point $u\in \CB$ of $\CG$ 
solving the equation (\ref{u}) and thus of $w':{B_\delta}\rightarrow 
h_{-n+1}$ given by Eq.\,\,(\ref{w'}) follows.
Using the natural embeddings (\ref{embedd}), we may consider
$B_n$ as a subset of ${B_\delta}$, and
$w'$ may be regarded as an element of the space $\CA_n$
of analytic functions from $B_n\subset h_n$ to $h_{-n}$.
Note also that, since 
$\tilde y=y+\Gamma w'(y)$ (see the argument of $\delta_2w$ 
in Eq.\,\,(\ref{u})), the inequality (\ref{11}) may 
be rewritten as
\qq
{\Vert}y +\Gamma w'(y){\Vert}_{_{n-1}} \leq \hf\m r^{n-1}\quad
\ {\rm for}\quad y\in {B_\delta}
\label{4.1}
\qqq
which implies that $y +\Gamma w'(y) \in B_{n-1}$ for such $y$.
\vskip 0.3cm

It is easy to see inductively that for $\vert\lambda\vert$ small 
enough, the maps $w_{n\beta}$ constructed in Proposition 2 give 
rise via the decomposition (\ref{w'}) to $u$'s which solve the fixed 
point equation (\ref{u}) and belong to $\CB$. It follows that $w'$ 
constructed above coincide for small $\vert\lambda\vert$, and 
hence for all $\lambda$ in the common domain of definition, 
with $w_{n\beta}$ of Proposition 2.
\vskip 0.4cm

\no{\bf Proof of  (b)}. \ Using the decomposition (\ref{w'}), 
we may write
\qq
w'(y)=w'(0)+Dw'(0)y+\delta_2w'(y),
\label{4}
\qqq
with 
\qq
w'(0)=Hw(0)+u(0)\m,\qquad Dw'(0)=HDw(0)+Du(0)
\label{6.1}
\qqq
and $\delta_2w'=\delta_2u$.
Let us first iterate the bound (\ref{w0}).
Note that 
\qq
P w'(0)=P H\m P w(0)+P u(0)
\nonumber
\qqq
since $H=H P$. As $u\in \CB$, see the definition (\ref{ball}),
we have,  
\qq
{\Vert}P u(0){\Vert}_{_{-n+1}}
\leq\,{\Vert}u(0){\Vert}_{_{-n+1}}
\leq\, 2\m\epsilon\m r^{{_3\over^2}(n-1)},
\label{A}
\qqq
which, with the use of the estimates (\ref{H2}) and (\ref{w0})
to bound ${\Vert}H\m P w(0){\Vert}_{_{-n+1}}$, implies that
\qq
{\Vert}P w'(0){\Vert}_{_{-n}}\leq\,
{\Vert}P w'(0){\Vert}_{_{-n+1}}
\leq\, 2\m\epsilon\, r^{2(n-1)}+ 2\m\epsilon\, 
r^{{_3\over^2}(n-1)}\m.
\label{bdw}
\qqq
\vskip0.2cm

This seems weaker than what is needed to iterate 
the estimate (\ref{w0}) but, as we shall see, one actually needs 
much less. The crucial point\footnote{This is the first 
of the two places where we gain by working with $w_{n\beta}$ 
for complex $\beta$.} is that the bound (\ref{bdw}) 
holds for $\;\vert{\rm Im}\m\beta\vert <\alpha_{n-1}$, while 
we have to establish the estimate (\ref{w0}) for $w'$ only for 
$\;\vert{\rm Im}\m\beta\vert<\alpha_{n}
=(1-n^{-2})\alpha_{n-1}$. For such $\beta$, 
we infer, using the estimate (\ref{bdw}), that, for $q\not=0$,
\qq
|w'(q;\m 0)|_{-n+1}\, e^{(\alpha_{n-1}-\alpha_{n})|q|}
\,\,\ee^{-\eta^{-n+1}\vert\omega
\cdot q\vert}\ \leq\ {\Vert}P w'(0){\Vert}_{_{-n+1}}\ 
\leq\ \epsilon
\label{bdw2}
\qqq
if in the middle term we take $w'$ corresponding 
to the value of $\beta$ shifted to $\beta'=\beta-i\m
\frac{(\alpha_{n-1}-\alpha_{n})}{\vert q\vert}\m q$. 
It follows that
\qq
\Vert P w'(q;\m 0)\Vert_{_{-n}}\ \leq\ \epsilon
\sum\limits_{q\not=0}\ee^{-n^{-2}\alpha_{n-1}\vert q\vert}
\,\,\ee^{-(1-\eta)\eta^{-n}\vert\omega\cdot q\vert}\m.
\label{omega.q}
\qqq
Since
$\alpha_{n-1}\geq\prod\limits_{n=2}^\infty(1-n^{-2})\,
\alpha>0$,   the sum on the right hand side
is clearly bounded  by $C n^2$. However, we may extract
from the sum factors that are super-exponentially small in $n$. 
Indeed, for $\vert\omega\cdot q\vert\leq \eta^{\hf n}$,  
we may extract from the first exponential under the sum 
a factor $\ee^{-\CO(\eta^{-{_n\over^{2\nu}}}n^{-2})}$ due 
to the Diophantine condition (\ref{Dio}). On the other hand, 
for $\vert\omega\cdot q\vert>\eta^{{\hf n}}$, we may extract 
a factor $\ee^{-\CO(\eta^{-\hf n})}$ from the second 
exponential. Hence the inductive bound 
(\ref{w0}) follows for $n\geq n_0$, and $n_0$ large enough.
\vs{4mm}

Let us now iterate the relation (\ref{omega}) 
for $\delta_2w'$ equal to $\delta_2u$  (see Eq.\,\,(\ref{w'})). 
Recall that ${\Vert}u(y){\Vert}_{_{-n+1}}
\leq 2\m\epsilon\m r^{{_3\over^2}(n-1)}$ 
for $\Vert y\Vert_{_{n-1}}<r^{n-\delta}$, see the definition
(\ref{ball}). The estimate (\ref{circ2}) with $k=2$ 
and $\gamma=r^{\delta}$ and the bounds (\ref{embe}) imply 
then that, for $\Vert y\Vert_{_n}<r^n\m$,
\qq
{\Vert}\delta_2w'(y){\Vert}_{_{-n}}\,\leq\,  
\frac{_{2\m r^{2\delta-{_3\over^2}}}}
{^{1-r^{\delta}}}\,\epsilon\, r^{{_3\over^2}n}\m.
\label{5}
\qqq
Taking $\delta>{3\over4}$ and $r<r(\delta)$, we infer that
${\Vert}\delta_2w'(y){\Vert}_{_{-n}}\leq\epsilon\, r^{{3\over2}n}$. 
This completes the inductive proof of (b).
\vskip 0.4cm

\no{\bf Proof of (c)}. We shall use the maps   $ \pi_{n\beta}$
introduced in Sect.\,\,5,
and related to $Dw_n$ by Eqs.\,\,(\ref{pioo}). Recall that   
$ \pi_{n\beta}$ was constructed as a map from $B(r^n)\subset h$ to 
$\CL (h,h)$, for $|\lambda|<\lambda_n$. With the use 
of the embeddings (\ref{embedd}), they may be viewed as
maps from $B_n \subset h_n$ to  $\CL (h_{n};h_{-n})$. 
As we shall see, they can be extended to  $|\lambda|<\lambda_0$.
Let us split $\pi_{n\beta}(\kappa;\m 0)$
into the diagonal and the off-diagonal parts:
\qq
\pi_{n\beta}(\kappa;\m 0)\,=\,\sigma_n(\kappa)+\rho_n(\kappa)\m,
\qqq
where $\sigma_n(q,q';\m\kappa)
=\pi_{n\beta}(q,q;\m\kappa;\m0)\,\delta_{q,q'}$.
Let us denote by $D_n$ the disc $\{\m\kappa\in\NC\,\,
\vert\,\,\vert\kappa\vert<\eta^{n}\m B\m\}$.
\vskip 0.6cm

\no{\bf Lemma 3}. 
{\it The maps $\,\pi\equiv\pi_{n\beta}: 
D_n\times B_n\rightarrow \CL(h_n;h_{-n})\,$ 
extend analytically to $\vert\lambda\vert<\lambda_0$.
Their extensions satisfy the relations 
\qq
t_p\m Dw(y)\,=\,\pi(\omega\cdot p\m;\, y)\ \ \quad{and}\ 
\ \quad t_p\m\pi(\kappa;\m y)\,
=\m\pi(\kappa+\omega\cdot p;\m y)\m,
\label{pi}
\qqq
whenever defined, and depend analytically on 
$\kappa$, $y$, $\beta$ and $\zeta$. They obey the bounds:
\qq
&&{\Vert}\m\delta_1\pi(\kappa ;\m y){\Vert}_{_{n;-n}}
\,\leq \,\epsilon\, r^{\ha n}\m ,
\label{pibound2}\\
&&\Vert\m\sigma(\kappa)\Vert_{_{n;-n}}
\,\leq\, \m\hf\m\epsilon\,\eta^{2n}\m ,
\label{taubound}\\
&&\Vert\m\rho(\kappa)\Vert_{_{n;-n}}
\,\leq\,\epsilon\, \m r^{{1\over 2}n}\m,
\label{rhobound}
\qqq 
where} $\delta_1\pi(\kappa;\m y)\equiv\pi(\kappa;\m y)
-\pi(\kappa;\m 0)$.
\vs{4mm}

\noindent Obviously, the bound (\ref{Dwbound}) follows by combining
Eqs.\,\,(\ref{pi}) for $p=0$, (\ref{pibound2}), (\ref{taubound}) 
and (\ref{rhobound}). In particular, we may use estimate 
(\ref{Dwbound}) as an inductive hypothesis in the proof
of Lemma 3. Hence, we are left with  
\vskip 0.6cm
 
\no{\bf Proof of Lemma 3.} \
We use the same notation as in the proof of Proposition 3.
Let us show that we may use recursively the relations 
(\ref{rinn}) to construct for $|\lambda|<\lambda_0$ the maps
$\pi_{n\beta}$ satisfying the identities (\ref{pi}). 
First, differentiating Eq.\,\,(\ref{wn+1}), we obtain
the relation
\qq
Dw'(y)\,=\,[\m1-Dw(\tilde y)
\m\Gamma\m]^{-1}Dw(\tilde y)\m,
\label{d'}
\qqq
where $\tilde y=y+\Gamma w'(y)$. The right hand side is well 
defined for $y\in {B_\delta}\subset h_{n-1}$, since, by inequality 
(\ref{4.1}), ${\tilde y}\in B_{n-1}$ for such $y$'s.
The bound (\ref{5.5}) and the inductive hypothesis (\ref{Dwbound})
imply that ${\Vert} Dw({\tilde y})\m\Gamma{\Vert}_{_{
-n+1;-n+1}} \leq\m C\m\epsilon$. Denoting 
$\pi\equiv\pi_{(n-1)\beta}$ we then define 
$\pi'\equiv\pi_{n\beta}$ by relation (\ref{rinn}), i.e. by
\qq
\pi'(\kappa;\m y)\ =\ [\m1-\,\pi(\kappa;\m{\tilde y})\,
\Gamma(\kappa)\m]^{-1}\m\pi(\kappa;\m {\tilde y})\m.
\label{1}
\qqq
The relations (\ref{pi}) for $\pi'$ may be inferred 
by applying the automorphism 
(\ref{auto}) to Eqs.\,\,(\ref{d'}) and (\ref{1}).  
The inductive hypotheses
imply that, for $\kappa\in D_{n-1}$ and $y\in {B_\delta}$, 
$\,\pi'(\kappa;\m y)$ is defined in $\CL(h_{n-1};h_{-n+1})$
and is an analytic function of its arguments. It follows 
by induction that it coincides for $\vert\lambda\vert<\lambda_n$
with the map $\pi_{n\beta}$ constructed before, 
see Eq.\,\,(\ref{pinn}), also satisfying the recursion 
(\ref{rinn}). Note that
\qq
\pi'(\kappa;\m 0)\ =\ 
[\m1-\,\pi(\kappa;\m{\tilde y_0})\,\Gamma(\kappa)\m]^{-1}
\m\pi(\kappa;\m {\tilde y_0})\m,
\label{111}
\qqq
where $\tilde y_0=\Gamma w'(0)$. The bounds 
\qq
&&{\Vert}\m\pi'(\kappa ;\m y)
{\Vert}_{_{n-1;-n+1}}
\m\ \ \ \leq\,\m \epsilon\,\eta^{2(n-1)}\m,
\label{brut0}\\
&&{\Vert}\m\delta_1\pi'(\kappa ;\m y)
{\Vert}_{_{n-1;-n+1}}
\,\leq\,\m \,3\,\epsilon\, r^{\hf(n-1)}
\label{brut}
\qqq
follow easily: the stronger factor $r^{\hf(n-1)}$ instead 
of the weaker one $\eta^{2(n-1)}$ in the estimate 
of $\delta_1\pi'$ is supplied by the inductive bound 
(\ref{pibound2}) for the difference
$\pi(\kappa;\m\tilde y)-\pi(\kappa;\m\tilde y_0)
=\delta_1\pi(\kappa;\m\tilde y)
-\delta_1\pi(\kappa;\m\tilde y_0)$. 
The inequality (\ref{brut}) is enough to iterate the bound 
(\ref{pibound2}). Indeed, the estimate (\ref{circ2}) 
with $k=1$ and $\gamma=r^{\delta}$ permits to extract 
the additional factor ${r^{\delta}\over 1-r^{\delta}}$
and to obtain the improved bound (\ref{pibound2}) 
(if $\delta>\hf$ and $r<r(\delta)$) for $\Vert\delta_1
\pi'(\kappa;\m y)\Vert_{_{n;-n}}$ for $y$ restricted 
to $B_n\subset h_n$.
\vskip 0.4cm

Let us turn to the inductive proof of the bound (\ref{rhobound})
for $\Vert\rho'(\kappa)\Vert_{_{n;-n}}$.
We gain a very small factor from the restriction of the analyticity 
strip in $\beta$, as before in the control of $P w'(0)$ 
in the proof of (b), see the inequalities (\ref{bdw2}). 
First, by the definition (\ref{norm}) of the norms, we have 
the following estimate of the kernel $\rho'(q,q';\m\kappa)$:
\qq
\vert\rho'(q,q';\m\kappa)\vert_{_{n-1;-n+1}}\,
\ee^{(\alpha_{n-1}-\alpha_{n})
\vert q-q'\vert}\,\m\ee^{-\eta^{-n+1}(\vert\omega\cdot q\vert
+\vert\omega\cdot q'\vert)}
\leq\m{\Vert}\m\pi'(\kappa;\m0)
{\Vert}_{_{n-1;-n+1}}\leq\, \epsilon\m,
\hspace{0.5cm}
\nonumber
\qqq
where in the middle term we have shifted $\beta$ to 
$\beta'=\beta-i\m\frac{(\alpha_{n-1}-\alpha_{n})}{\vert q-q'\vert}
\m(q-q')$. The last inequality follows from
 the bound (\ref{brut0}). 
Moreover, for an operator with  kernel 
$a(q,q')$, $\,\Vert a\Vert_{_{n;-n}}\,\leq\,
\sup\limits_{q'}\sum\limits_{q}\,\vert a(q,q')\vert_{_{n;-n}}
\,\m\ee^{-\eta^{-n}(\vert\omega\cdot q\vert
+\vert\omega\cdot q'\vert)}\m.\,$ This implies the estimate
\qq
{\Vert}\m\rho'(\kappa){\Vert}_{_{n;-n}}\,
\leq\, \epsilon\,\m\sup\limits_{q'}\sum\limits_{q\not=q'}
\ee^{-n^{-2}\m\alpha_{n-1}\vert q-q'\vert}
\,\,\ee^{-(1-\eta)\eta^{-n}(\vert\omega\cdot q\vert
+\vert\omega\cdot q'\vert)}
\nonumber
\qqq
The expression on the right hand side may be bounded by a factor
super-exponentially small in $n$, which is much more than is needed.
Indeed, we may extract from it an extra factor 
$\ee^{-\CO(\eta^{-{n\over 2\nu}}n^{-2})}$ 
for $\vert\omega\cdot q\vert$ and $\vert\omega\cdot q'
\vert\leq\eta^{\hf n}$ and hence $0\not=\vert\omega
\cdot(q-q')\vert\leq 2\eta^{\hf n}$ using the Diophantine condition 
(\ref{Dio}) and an extra factor $\ee^{-\CO(\eta^{-\hf n})}$
if $\vert\omega\cdot q\vert>\eta^{\hf n}$ or 
$\vert\omega\cdot q'\vert>\eta^{\hf n}$. Hence 
the bound (\ref{rhobound}) for the off-diagonal operator 
$\rho'(\kappa)$.
\vskip 0.4cm

We are left with the proof of the estimate
(\ref{taubound}) for the diagonal operator $\sigma'(\kappa)$.
Let us define
\qq
{s}(z)\,=\, u^{n-1}\m\sigma(0,0;B\eta^{n-1}z)\, u^{n-1}\m,
\label{tild}
\qqq
where $u=(\matrix{_1&_0\cr^0
&^\eta })$ is a block matrix. Similarly, we introduce 
the matrix $s'(z)$ related to $\sigma'$. 
With the use of symmetry (\ref{sz1}), we write 
$s'(z)=\eta^{2(n+1)}(\matrix{_{\wp'_0(z)}
&_{\wp'_1(z)}\cr^{\wp'_1(-z)}&^{\wp'_2(z)} })$. 
We shall prove that, for $|z|<1 $,
\qq
&&|\wp'_i(z)-p'_i\, z^{2-i}|\,\m\leq\ A\m |z|^{3-i}
\label{taub0}
\qqq
with $|p'_i|\leq (1-{1\over{n}}){\epsilon\over{32}}$ and 
$A\leq {\epsilon\over32}$, assuming inductively similar bounds 
for $s(z)$. Note that such inductive assumptions, together 
with the identity $|M|_{_{n-1;-n+1}}=|u^{n-1}Mu^{n-1}|_{_{0;0}}$ 
for the matrix norms imply, in particular, the estimate
\qq
|\sigma(0,0,\kappa)|_{n-1;-n+1} \leq {_1\over^8}\m\epsilon
\,\eta^{2n}
\label{smallk}
\qqq
for $\kappa\in D_{n-1}$.
The bound (\ref{taub0})  will follow from
Lemma 2 expressing the cancellations of resonances. 
The leading Taylor coefficients $p_i$ of $\wp_i(z)$ 
are {\it marginal} in the RG terminology and the higher ones  
are {\it irrelevant}. The presence of lower order {\it relevant} 
Taylor coefficients would spoil the iterative bounds.
They are, however, forbidden by the Ward identities.
Let us pass to the details.
\vs{4mm}

Let us first prove the estimate (\ref{taubound}) for $\sigma'$
assuming the bound (\ref{smallk}). We shall split 
\qq
\sigma'(\kappa)\,
=\,\sigma'_0(\kappa)+\sigma'_1(\kappa)\qquad{\rm with}\qquad
\sigma'_0(\kappa)\ =\ [\m1-\,\sigma(\kappa)\m
\Gamma(\kappa)\m]^{-1}\m\sigma(\kappa)\m,
\label{112}
\qqq
compare with  Eq.\,\,(\ref{111}). 
Note that $\sigma(\kappa)\m\Gamma(\kappa)$ is an
operator diagonal in  Fourier space and hence, so is 
$\sigma'_0(\kappa)$. Since, by the inductive hypotheses
(\ref{rhobound}) and (\ref{pibound2}),
\qq
\Vert\sigma(\kappa)-\pi(\kappa;\m\tilde y_0)
\Vert_{_{n-1;-n+1}}\,=\,\Vert\rho(\kappa)
+\delta_1\pi(\kappa;\m\tilde y_0)
\Vert_{_{n-1;-n+1}}\,\leq\, 
2\m\epsilon\, r^{{1\over 2}(n-1)}\m,
\label{prnpr1}
\qqq
it follows that
\qq
\Vert\m\sigma'_1(\kappa)\Vert_{_{n;-n}}
\,\leq\,{_1\over^4}\m \m\epsilon\,\eta^{2n}\m,
\label{prnpr}
\qqq
for $r$ small enough.
We pass to the estimation of $\sigma'_0(\kappa)$.
Note that the bound (\ref{5.5}) together with the 
inequalities (\ref{embe}) and the inductive
hypothesis (\ref{taubound}) imply that 
$\Vert\Gamma(\kappa) \m\sigma(\kappa)
\Vert_{_{-n;-n}}\leq\,\hf
\m C\m\epsilon$ so that
\qq
\Vert\sigma'_0(\kappa)\Vert_{_{n;-n}}\leq\, 
2\m\Vert\sigma(\kappa)
\Vert_{_{n;-n}}\m.
\label{tt0}
\qqq
For operators $a$ diagonal in Fourier transform,
$\,{\Vert} a{\Vert}_{_{n;-n}}
=\sup\limits_q\vert a(q,q)\vert_{_{n;-n}}\,\m\ee^{-2\eta^{-n}\vert
\omega\cdot q\vert}\m.\,$
Hence it follows from the bound (\ref{tt0}) that
\qq
\Vert\sigma'_0(\kappa)\Vert_{_{n;-n}}\ \leq\ 
\sup\limits_q\ 
2\,\vert\m\sigma(q,q;\m\kappa)\m\vert_{_{n;-n}}
\m\,\ee^{-2\eta^{-n}\vert\omega\cdot q\vert}\m.
\label{esse}
\qqq
For $q$ with $\vert\omega\cdot q\vert<(1-\eta)\eta^{n-1}B$,
we use for $\kappa\in D_n$ the equality
$\sigma(q,q;\m\kappa)=\sigma(0,0;\m\tilde\kappa)$
with $\tilde\kappa=\kappa+\omega\cdot q$
which follows from the second identity (\ref{pi})
(observe that for such $q$'s and for $\kappa\in D_n$,
${\tilde \kappa}\in D_{n-1}$).
By virtue of the inequality (\ref{smallk}),
\qq
|\sigma(0,0;\tilde\kappa)|_{_{n;-n}}
\leq\,|\sigma(0,0;\tilde\kappa)|_{_{n-1;-n+1}}
\leq\,{_1\over^8}\m \m\epsilon\,\eta^{2n}\m.
\hspace{0.3cm}
\qqq
Hence, for $q$ with $\vert\omega\cdot 
q\vert<(1-\eta)\eta^{n-1}B$,
we may bound the expression on the right hand side 
in the estimate (\ref{esse}) by ${_1\over^4} \epsilon\m
\eta^{2n}$. For $\vert\omega\cdot q\vert\geq(1-\eta)
\eta^{n-1}B$, we instead extract an extra factor 
estimating
\qq
2\,\vert\m\sigma(q,q;\m\kappa)\m\vert_{_{n;-n}}
\,\ee^{-2\eta^{-n}\vert\omega\cdot q\vert}\,\leq\,
2\,\Vert\sigma(\kappa)\Vert_{_{n-1;-n+1}}
\,\m\ee^{-2\eta^{-1}(1-\eta)^2 B}\m\cr\cr
\leq\,\ee^{-2\eta^{-1}(1-\eta)^2 B}\, \m\epsilon\m
\eta^{2(n-1)}\,\leq\,{_1\over^4}\m\epsilon\m\eta^{2n}
\label{jb2}
\qqq
for $B$ sufficiently large (this is the only place where
$B$ large is needed). Putting these estimates together 
with the inequality (\ref{prnpr}) for $\sigma'_1$,
we infer the bound (\ref{taubound}) for $\sigma'(\kappa)$.
\vskip 0.4cm

We still have to iterate the crucial estimates (\ref{taub0})
which is the only place in the proof of Proposition 3 where 
we use the Ward identities. 
Writing  Eqs.\,\,(\ref{112}) in terms of $s$, see the definition
(\ref{tild}), we obtain
\qq
s'(z)\,=\,[1-(\CL s)(z){\tilde\gamma}(z)]^{-1}(\CL s)(z)\,
+\, s'_1(z)
\nonumber
\qqq
with the ``linearized RG map'' $\CL$, 
$$
(\CL s)(z)=u\m s(\eta z)\m u\m,
$$
and ${\tilde\gamma}(z)=u^{-n}\gamma_{n-1}(B\eta^nz)u^{-n}=
\eta^{-2n}(Bz)^{-2}\chi_1(B\eta^2 z)
(\matrix{_\mu &_{iBz}\cr ^{-iBz} & ^0})$, \,see Eqs.\,\,(\ref{chi}) 
and (\ref{gamma}). The estimate (\ref{prnpr1}) implies that
the remainder $s'_1$ satisfies the bound 
$|s'_1(z)|_{_{0,0}}\leq C\m\epsilon\, r^{{1\over 2}n}$. 
Combining the definition (\ref{chi0}) (which implies that  $\chi_1(z)$
is of order $|z|^6$ for small $z$) and the inductive  bound for $s$,
we infer that $|(\CL s)(z){\tilde\gamma}
(z)|_{_{0;0}}\leq C\m\epsilon\m|z|^4$. Thus altogether
$$
|s'(z)-(\CL s)(z)|_{_{0;0}}\m\leq\,\m  C\epsilon^2\eta^{2n} |z|^4
+C\m\epsilon\, r^{\hf n}\m.
$$
The map $\CL$ preserves $p_i$ and contracts the constant $A$ to
$\eta\m A$. The Ward identity, Lemma 2, implies that 
$\partial^j\wp'_i(0)=0$ for $j<2-i$. Since 
$\vert p'_i-p_i\vert\leq C\m\epsilon\,\eta^{-2n}\m r^{\hf n}
\leq{\epsilon\over32 n^2}$ and $\eta\m A+C\m\epsilon^2
+C\m\epsilon\m\eta^{-2n}\m r^{\hf n}\leq A$ \,for $r$ 
and $\epsilon$ small, we infer that $s'$ satisfies the bound 
(\ref{taub0}). This finishes the proof of Lemma 3 
and of Proposition 3.
\ \ $\Box$
\vskip 0.2cm

\nsection{ Proof of Theorem 1}  
\vskip 0.2cm

We shall first show that $X_n\equiv F_n(0)=\Gamma_{<n}W_n(0)$
converges to a real analytic function $X$ with zero average
as $n\to \infty$ and that $X$ solves Eq.\,\,(\ref{fp1}). 
\vskip 0.4cm

Recall that in  Proposition 2, we have constructed
for $\vert\lambda\vert<\lambda_n$ the analytic maps 
$f_{n\beta}$ from $B(r^n)\subset h$ into $h$, satisfying 
the relations (\ref{fnc}) and (\ref{fn+1}) and the bound 
$\vert\vert\vert f_{n\beta}\vert\vert\vert\leq 2\m r^n$.
They may be also viewed as analytic maps from
$B(r^n)\subset h_n$ to $h$. As such, they may be analytically
extended to $\vert\lambda\vert<\lambda_0$ for $n\geq n_0$
by iterated use of Eq.\,\,(\ref{fn+1}) if we recall the bound 
(\ref{4.1}). The extensions are clearly uniformly bounded
(e.g. by $2\m r^{n_0}$). 
Let us prove the convergence in $h$ of $x_{n\beta}
\equiv f_{n\beta}(0)$ obtained this way. The recursion (\ref{fn+1}) 
implies that
\qq
x_{n\beta}\,=\, x_{(n-1)\beta}\m
+\m\delta_1 f_{(n-1)\beta}(\Gamma_{n-1} w_{n\beta}(0))\m.
\label{cnv}
\qqq
Using the estimate (\ref{circ2}) for $k=1$,
we deduce from the inequalities (\ref{5.5}) and (\ref{w0})
that the $\Vert\cdot\Vert$ norm of the 2$^{\rm nd}$ term 
on the right hand side of Eq.\,\,(\ref{cnv}) is bounded  
by $C\m\epsilon\, \eta^{-2n}\m r^{2n}$. Hence the convergence
\qq
x_{n\beta}\ \ \rightarrow\ \ x_\beta
\qquad{\rm in}\ \quad h
\label{cnv1}
\qqq
together with the bound $\Vert x_\beta\Vert\leq 
C\m\epsilon$
uniform in the strip 
$\vert{\rm Im}\m \beta\vert<\alpha'=\alpha \prod_{n=2}^\infty 
(1-n^{-2})$. The latter implies the pointwise estimate
\qq
\vert x(q)\vert\ \leq\  C\m\epsilon\,
\ee^{-\alpha'\m\vert q\vert}
\label{man}
\qqq
and, consequently, the real analyticity of the Fourier transform 
$X$ of $x$.
\vskip 0.4cm

For $\vert\lambda\vert<\lambda_n$,
Eqs.\,\,(\ref{fnc}) and (\ref{wnc})
imply that
\qq
x_{n}\equiv f_{n}(0)\m=\m\Gamma_{<n}\m w_0(x_n)
\qquad{\rm and}\qquad w_0(x_{n})\m=\m w_n(0)\m.
\label{smll}
\qqq
In particular, it follows from the first of these equations
that $x_n(q)\vert_{_{q=0}}=0$  and from the second one 
and Lemma 1 (i.e.\,\,the Ward identities) 
that $w_0^i(q;\m x_{n})\vert_{_{q=0}}=0$ 
for $i\leq d$. By analyticity in $\lambda$, these relations 
have to hold for $\vert\lambda\vert\leq\lambda_0$.
Since $w_0$ is analytic, we can take 
the $n\to \infty$ limit of Eqs.\,\,(\ref{smll}), 
and infer that
\qq
x(0)\m=\m0\m,\quad x(q)\m=\m G_0\m w_{0}(q;\m x)\;\;
{\rm for}\quad q\not=0\m, \;\;{\rm and}\;\;
w_0(0;\m x)\,=\,(0,\xi)\m.
\label{3eq}
\qqq
The first 2 of these equations 
are the Fourier transformed
version of Eq.\,\,(\ref{fp1}). 
The solution $x$ is an analytic function of $\lambda$
for $\vert\lambda\vert<\lambda_0$ and it vanishes
for $\lambda=0$. Recall that $\zeta$ is a parameter 
in $w_0$, and thus $x$ is also analytic in $\zeta$ 
for  $|\zeta|<r_0$. 
\vskip 0.4cm

We still have to solve Eq.\,\,(\ref{fp12}) for $\zeta$.
In view of the $3^{\rm rd}$ of Eqs.\,\,(\ref{3eq}) and
Eq.\,\,(\ref{W1}) it reduces to the equalities
\qq
\mu\zeta\,+\,\lambda\int_{_{\NT^d}}\partial_I U((\phi,\zeta)
+X(\phi))\,\m d\phi\,=\,0\ \quad{\rm or}\ \quad\int_{_{\NT^d}}
\partial_I U((\phi,\zeta)+X(\phi))\,\m d\phi\,=\,0
\qqq
in the, respectively, non-isochronous and isochronous cases.  
We shall solve the above equations for $\zeta$ by the Implicit
Function Theorem. Note that $\lambda=0$ and $\zeta=0$ satisfies 
the non-isochronous equation  and that the $\zeta$-derivative
of its left hand side at these points is $\mu$, which is assumed to be
invertible. Similarly, $\lambda=0$ and $\zeta=0$ solves
the isochronous equation (since $X$ vanishes for $\lambda=0$ 
and we have assumed that $\int\partial_I U(\phi,0)\,d\phi=0$) 
and the $\zeta$-derivative of its left hand side at these points 
is $\int\partial_I^2U(\phi,0)\, d\phi$, which is assumed 
to be invertible. 
The existence of the local solution $\zeta(\lambda)$ analytic 
for $\vert\lambda\vert<\lambda_0$ with $\lambda_0$ small enough 
and vanishing at $\lambda=0$ follows in the both cases. 
The resulting solution $Z=X+(0,\zeta)$ of Eq.\,\,(\ref{Z}) 
depends analytically on $\lambda$ for $\vert\lambda\vert<\lambda_0$ 
and vanishes for $\lambda=0$. Its uniqueness up to translations 
(\ref{Ztr}) follows from the fact that the equations (\ref{fp1}) 
and (\ref{fp12}) completely determine the coefficients 
of the Taylor expansion of their solution in powers of $\lambda$
(i.e.\,\,the Lindstedt series discussed in Sect.\,\,9). 
This ends the proof of Theorem 1. \ \ $\Box$
\vskip 0.2cm

\nsection{Field theory interpretation}
\vskip 0.2cm

\noindent Although arising in a problem of classical mechanics
of $d$ degrees of freedom, Eq.\,\,(\ref{Z}) has a field theory 
interpretation, see \cite{GalPar,GGM}. Let us consider the (action) 
functional $S(Z)$ of maps $Z=(\Theta,J):\NT^d\rightarrow
\NR^d\times\NR^d$,
\qq
S(Z)\ =\ \int\limits_{\NT^d}\left[\m{_1\over^2}\m J(\phi)\cdot
\mu\m J(\phi)-J(\phi)\cdot(\omega\cdot\da_\phi)\Theta(\phi)
+\lambda  
U(\phi+\Theta(\phi),J(\phi))\m\right]\, d\phi\m.
\label{func}
\qqq
Note the translation symmetry
\qq
S(\tau_\beta Z\m-\m(\beta,0))\ =\ S(Z)\m,
\label{sym1}
\qqq
where, as before, $(\tau_\beta Z)(\phi)=Z(\phi-\beta)$
for $\beta\in\NR^d$. As is easily seen, our basic equation 
(\ref{Z}) coincides with the equation $\delta S(Z)=0$ for 
the extrema of functional $S$. The map $Z$ with 
$\int\Theta(\phi)\,d\phi=0$ minimizing $S$ may, in turn, 
be interpreted as the limit $\hslash\rightarrow 0$ 
of the formal functional integral 
\qq
\Big\langle Z(\phi)\Big\rangle_{\hslash}
\ =\ {\int Z(\phi)\ \ee^{\m{i\over\hslash}\m S(Z)} 
DZ\over\int\ee^{\m{i\over\hslash}\m S(Z)}\,\, DZ\m,}
\label{FI1}
\qqq
where $DZ$ denotes the formal Lebesgue measure 
on $L_0^2(\NT^d,\NR^d)/\NZ^d\times L^2(\NT^d,\NR^d)$ 
where $L^2_0$ is composed of maps with zero average.
Indeed, in this limit, the integral should be localized 
at the minimum of the functional (\ref{func}).
\vskip 0.4cm

While Eq.\,\,(\ref{FI1}) has a purely formal meaning, we may
gain intuition from it by some further manipulations.
First,  write $Z=(0,\zeta)+Y$ with $Y$ of zero average. 
Then  
\qq
\Big\langle Z(\phi)\Big\rangle_{\hslash}=
\ {\int\m [\m(0,\zeta)+Y(\phi)]\ \ee^{\m{i\over\hslash}\, 
V_0(Y,\zeta)}\,\, d\mu_0(Y)\m d\mu(\zeta)\over 
\int\ee^{\m{i\over\hslash}\,V_0(Y,\zeta)}\,\,d\mu_0(Y)\m 
d\mu(\zeta)}\m,
\label{FI2}
\qqq
where
\qq
V_0(Y,\zeta)=\lambda \int\limits_{\NT^d} 
U((\phi,\zeta)+Y(\phi))d\phi 
\label{V1}
\qqq
and we have used the quadratic part of the action functional 
to define the oscillatory ``measures''
\qq
d\mu(\zeta)\ =(\det\m{_\mu\over^{2\pi i\hslash}})^\hf\,\m 
\ee^{\m{i\over2\m\hslash}\,\zeta\cdot\mu\zeta}\,\, d\zeta\
\label{mugi}
\qqq
and $d\mu_0$. The latter is formally given by
\qq
d\mu_0(Y)\ =\ \ee^{-{i\over2\m\hslash}\,
(Y,\m G_0^{-1}Y)}\,\, DY\ \Bigg/\ 
\int\ee^{-{i\over2\m\hslash}\,
(Y,\m G_0^{-1}Y)}\,\, DY\m,
\label{mug}
\qqq
i.e. it is the ``Gaussian measure'' with mean zero and 
covariance $i{\hslash}G_0$ on the space of maps $Y$ with zero 
average.
\vs{2mm}

In quantum field theory, the idea of the renormalization group (RG) 
is to calculate functional integrals inductively. Let us
explain a concrete realization of this idea. Let
\qq
G_0=G_1+\Gamma_0
\nonumber
\qqq
be a decomposition of the operator $G_0$ into a sum of two
operators. We may then write  the measure 
$d\mu_0$ as a product measure according to the formula:
\qq
\int F(Y)\,\, d\mu_0(Y)\ =\ \int F(Y+\m\tilde Y)
\,\, d\mu_1(Y)\,\, d\nu_0(\tilde Y)\m,
\label{prod}
\qqq
where $d\mu_1$ is the measure given by Eq.\,\,(\ref{mug})
with $G_0$ replaced by $G_1$ and $d\nu_0$ is the Gaussian
measure with mean zero and covariance $i{\hslash}\Gamma_0$, 
both on the space of maps with zero average.
Using this identity, we may rewrite Eq.\,\,(\ref{FI2}) as
\qq
\Big\langle Z(\phi)\Big\rangle_{\hslash}=
{\int\m [\m(0,\zeta)+ F_1(Y,\zeta;\phi)]
\ \ee^{\m{i\over\hslash}\, V_1(Y,\zeta)}
\,\, d\mu_1(Y)\m d\mu(\zeta)\over \int\ee^{\m{i\over\hslash}\, 
V_1(Y,\zeta)}\,\, d\mu_1(Y)\m d\mu(\zeta)}\,
\label{FI3}
\qqq
if we set
\qq
&&\ee^{\m{i\over\hslash}\, V_1(Y,\zeta)}\ =\ 
\int\ee^{\m{i\over\hslash}\, V_0(Y+\tilde Y,\m
\zeta)}\,\, d\nu_0(\tilde Y)\,,
\label{effi}\\
&&F_1(Y,\zeta;\phi)\ =\  
{\int\m [\m Y(\phi)+\tilde Y(\phi)\m]\,\,\ee^{\m{i\over\hslash}\, 
V_0(Y+\tilde Y,\m\zeta)}\,\, d\nu_0(\tilde Y)\over
\int\ee^{\m{i\over\hslash}\, V_0(Y+\tilde Y,\m\zeta)}
\,\, d\nu_0(\tilde Y)}\m.
\label{effe}
\qqq
This is the first step of the iterative procedure. After 
the $n$ subsequent decompositions, 
\qq
G_0\ =\ G_{n}\,+\,\sum\limits_{k=0}^{n-1}\Gamma_k\m,
\nonumber
\qqq
one arrives at the expression
\qq
\Big\langle Z(\phi)\Big\rangle_{\hslash}=
{\int\m [\m (0,\zeta)+ F_{n}(Y,\zeta;\phi)\m]
\ \ee^{\m{i\over\hslash}\, V_{n}(Y,\zeta)}
\,\, d\mu_{n}(Y)\m d\mu(\zeta)\over \int\ee^{\m{i\over\hslash}\, 
V_{n}(Y,\zeta)}\,\, d\mu_{n}(Y)\m d\mu(\zeta)}
\label{FIn}
\qqq
with
\qq
&&\ee^{\m{i\over\hslash}\, V_{n}(Y,\zeta)}\ =\ 
\int\ee^{\m{i\over\hslash}\, V_{n-1}(Y+\tilde Y,\m\zeta)}
\,\, d\nu_{n-1}(\tilde Y)\,,
\label{effin}\\
&&F_{n}(Y,\zeta;\phi)\ =\  
{\int F_{n-1}(Y+\tilde Y,\m\zeta;\phi)\,\,\ee^{\m{i\over\hslash}
\, V_{n-1}(Y+\tilde Y,\m\zeta)}
\,\, d\nu_{n-1}(\tilde Y)\over
\int\ee^{\m{i\over\hslash}\, V_{n-1}(Y+\tilde Y,\m\zeta)}\,\, 
d\nu_{n-1}(\tilde Y)}\m.
\label{effen}
\qqq
Such a procedure may lead to the calculation of the expectation
$\,\langle\m Z(\phi)\m\rangle_{_{\hslash}}\,$ if we
have a good control of the asymptotic behavior of the 
effective interactions $V_n$ and of the effective insertions
$F_n$. Note that the translation symmetry (\ref{sym1})
goes through the RG iteration (\ref{effin}): 
\qq
V_n(\tau_\beta Y\m-\m(\beta,0))\ =\ V_n(Y)
\label{symn}
\qqq
if we consider in the definitions (\ref{effi}) and (\ref{effin})
fields $Y$ with arbitrary averages and if the covariances $\Gamma_n$ 
commute with the translations $\tau_\beta$. The latter property has 
been guaranteed by the explicit construction of the operators 
$\Gamma_n$ (see Eq.\,\,(\ref{Gamma})).
\vskip 0.5cm

The inductive RG scheme described in Sect.\,\,2 may be obtained as
the formal $\hslash\rightarrow 0$ limit  of the above iterative
calculation of the functional integral (\ref{FI1}). 
After the first step, we obtain
\qq
V_1(Y)\ =\ -\hf\,(\tilde Y_0,\m\Gamma_0^{-1}\tilde Y_0)
\,+\, V_0(Y+\tilde Y_0)\,,\qquad F_1(Y)\ =\ Y+\tilde Y_0\m,
\label{infty}
\qqq
where $\tilde Y_0$ minimizes the right hand side of the first
equation. Denoting 
\qq
W_n(\phi;Y)\ =\ {{\delta V_n(Y)}\over{\delta Y(\phi)}}\,,
\label{W}
\qqq
we may rewrite Eqs.\,\,(\ref{infty}) in the differential 
form:
\qq
W_1(Y)\ =\ W_0(Y+\tilde Y_0)\,,\qquad
\ \tilde Y_0\ =\ \Gamma_0\m W_0(Y+\tilde Y_0)
\qqq
or, more conveniently, as the fixed point equation
\qq
W_1(Y)\ =\ W_0(Y\m+\m\Gamma_0 W_1(Y))
\label{fP}
\qqq
whose solution determines $F_1$:
\qq
F_1(Y)\ =\ Y+\tilde Y_0\m  =\ Y\m+\m\Gamma_0\m W_1(Y)\m.
\label{F1}
\qqq
These are Eqs.\,\,(\ref{W2}) and (\ref{F2}), respectively.
Similarly, after $n$ inductive steps, $W_{n}$ 
and $F_{n}$ are determined by relations (\ref{Wn+1}) 
and (\ref{Fn+1}), the $\hslash\rightarrow 0$ versions 
of Eqs.\,\,(\ref{effin}) and (\ref{effen}), respectively. 
The cumulative expressions (\ref{cum1}) 
and (\ref{cum2}) may be obtained as those for $W_1$ and $F_1$ 
by replacing $\Gamma_0$ by $\Gamma_{<n}=\sum\limits_{k<n}
\Gamma_k$, \, i.e.{\,\,}by performing $n$ RG steps at once.
\vskip 0.4cm

The Ward identity (\ref{ward}) of Sect.\,\,5, which assured 
the partial cancellation of the repeated resonances or,
in the field theory language, the absence of the terms
proportional to $\,\int|\Theta|^2$, $\,\int\Theta\cdot
(\omega\cdot\partial)\Theta$, $\,\int J\cdot\Theta$ (and
to $\,\int\Theta$) in the effective interactions $V_n$, is 
the infinitesimal version of the translation 
symmetry (\ref{symn}). It is obtained from the latter by 
the differentiation w.r.t. $\beta^i$ at $\beta=0$, which yields
\qq
-\int{\delta V_n(Y)\over\delta Y^i(\phi)}\,\, d\phi
\ -\ \sum\limits_l\int {\delta V_n(Y)\over\delta Y^l(\phi)}
\,\,\partial_{\phi^i} Y^l(\phi)\,\, d\phi\ =\ 0\m.
\nonumber
\qqq
The integration by parts in the second term and the definition
(\ref{W}) give then the identity (\ref{ward}).

\nsection{Renormalization Group and Lindstedt series.}
\vs{2mm}

In this section, we sketch the
connection of our approach to the Lindstedt series
and to the resummation of the latter by Eliasson. 
The Lindstedt series has a graphical representation 
and our RG method amounts to resumming
at each step a particular subset of graphs.
Let us first introduce  the graphical representation. 
It will be convenient to use a common symbol $Q$ 
for the momentum variable $q$ and the vector index $i$, 
$Q=(q,i)\in\NZ^d\times\{1,\dots,2d\}\equiv
\CI$ and to write $x(Q)\equiv x^{i} (q)$ and 
$w_0^{(m)}(Q_0 \dots,Q_m)$ for the matrix
elements of the kernels  $w_0^{(m)}(q_0, q_1 \dots,q_m;\zeta)$
introduced in Eq.\,\,(\ref{taylor}). Using the relations 
(\ref{fp1}) and (\ref{taylor}),  we obtain
\qq
x(Q)\,=\sum\limits_{m=0}^{\infty}
{\sum\limits_{{\bf Q}}}^*G_0(Q , Q_0)\,
\, w_0^{(m)}({\bf Q})
\prod_{\ell=1}^m x({Q_\ell})\m,
\label{wkernel0}
\qqq
where 
$G_0(Q, Q_0)\equiv G_0(q , q_0)^{i i_0}$ 
and ${\bf Q}= (Q_0, \dots Q_m)$. 
The sum $\sum^*$ means that we sum over $q_i\neq 0$
(this is due to the projector $P$ in Eq.\,\,(\ref{fp1})).
Since $w_0^{(m)}$ is proportional to $\lambda$, this formula
yields a power series solution in powers of $\lambda$ of $x(Q)$
obtained by regarding the equality (\ref{wkernel0}) as a fixed 
point equation solved iteratively, starting with $x=0$.
\vskip 0.3cm

The resulting series can be conveniently expressed as
a sum over tree graphs whose weights are as follows.
Let $\CT_{m+1}^k$ denote the set of  connected
tree graphs $T$ on $m+1+k$ vertices $v\in\{0,\dots,m+k\}\equiv 
V(T)$ such that the first $m+1$ ones, the ``external
vertices'', have  coordination number one. We shall
assume that $m+1,k\geq 1$ and will call the first of the external
vertices the root of $T$. The lines $\ell$ of $T$ are pairs 
$\ell=(v,v')$, $v,v'\in V(T)$ which we order assuming 
that the unique path going from the root to 
$v'$ goes through $v$. 
The set of lines is denoted by $\CL(T)$, lines $\ell$ containing
an external vertex are called external, $\ell\in\CL_E(T)$, and 
the  remaining ones internal, $\ell\in\CL_I(T)$.
Given a function $G:\CI^2\rightarrow\NC$ and a collection
$(w^{(m)})_{m\geq 0}\equiv\, w$ of functions 
$w^{(m)}:\CI^{m+1}\rightarrow\NC$ symmetric 
in the last $m$-variables, we define the ``amplitude'' 
$A(T,G,w,\NQ)$ of $T$. For this purpose, we assign variables 
$Q_i\in \CI$ to the external vertices $0\leq i\leq m$ of $T$
and variables $P_{\ell v}\in \CI$ to each internal line $\ell$ 
and a vertex $v$ contained in it. We write $\NP_\ell
=(P_{\ell v},P_{\ell v'})$ for $\ell = (v,v')$. For 
$m<v\leq m+k$, we set $R_v=P_{\ell v}$ if there exists 
an internal line $(v',v)$ or $R_v=Q_0$ otherwise 
(i.e. when $(0,v)$ is an external line), and 
$\NP_v=\{\m P_{\ell v}\,|\,\ell\in\CL_I(T),\ \ell=(v,v')\m\}$,
$\NQ_v=\{\m Q_i\,|\,\ell=(v,i)\in\CL_E(T)\m\}$. 
We define:
\qq
A(T,G,w,\NQ)\,=\,{_1\over^{m!\, k!}}\,{\sum_{\NP}}^*
\prod_{\ell\in\CL_I(T)}G(\NP_\ell)\prod_{v=m+1}^{m+k} m_v!
\m\, w^{(m_v)}(R_v,\NP_v,\NQ_v)\m,
\label{w}
\qqq
where $\NQ=(Q_0,\dots,Q_m)$, the sum ${\sum\limits_{\NP}}^*$ 
runs over all $p_{\ell v} \neq 0$, and $m_v +1$ is the coordination 
number of vertex $v$.
\vskip 0.4cm

With this notation, the iterative solution 
of Eq.\,\,(\ref{wkernel0}) is given by
\qq
x(Q)={\sum_{Q_0}}^*\sum_{T\in\CT_1}G_0(Q,Q_0)\,\m 
A(T,G_0,w_0,Q_0)\m,
\label{sol}
\qqq
where the sum is over all tree graphs with one external
line, $\CT_1=\cup_k\CT_1^k$.
Comparing with Eq.\,\,(\ref{wkernel0}), we also infer that
\qq
w_0 (Q_0;x) =
\sum_{T\in\CT_1} A(T,G_0,w_0,Q_0)
\label{wcl}
\qqq
\vs{4mm}

\no {\bf Remark.} \ A formal way to derive the identity
(\ref{sol}) is to start from the field theory formula (\ref{FI2}),
expand it in Feynman diagrams, and let $\hslash \to 0$. In that
limit, only tree diagrams remain (each line has a power
of $\hslash $, while each vertex carries a factor $\hslash ^{-1}$;
so, all graphs except tree graphs are multiplied by some positive power
of $\hslash $), and we obtain Eq.\,\,(\ref{sol}).
\vs{4mm}

As is well known since Poincar\'e \cite{Po}, the series 
in Eq.\,\,(\ref{sol}) does not absolutely converge for nontrivial 
potentials $U$ in the Hamiltonian (\ref{H}), i.e.
$\sum_{T\in\CT_1}|A(T,G_0,w_0,Q_0)|=\infty$ (see e.g. \cite{CF1} 
for a simple proof). This is due to the presence of repeated 
resonances, namely  of long sequences of lines (connected by
vertices with coordination number two)
many of which have the same small denominators
$\omega \cdot p_\ell$. However, many trees 
contribute to the same order in $\lambda$ and,
as shown by Eliasson \cite{E}, cancellations
occur when one regroups terms in a suitable way.
\vskip 0.4cm

Let us now explain the renormalization group in the graph 
language. The key idea is a combinatorial identity which
performs a partial resummation in Eq.\,\,(\ref{sol}):
\vs{5mm}

\no{\bf Proposition 4}. {\it Let $\m G=G'+\Gamma$. Then,
\qq
\sum_{T\in\CT_{m+1}}A(T,G,w,\NQ)=\sum_{T'\in\CT_{m+1}}A(T',G',w',\NQ)
\label{rg}
\qqq
with}
\qq
w^{'(m)}(\NP)=\sum_{T\in\CT_{m+1}}A(T,\Gamma,w,\NP)
\label{rg1}
\qqq
\vs{5mm}

The   proof of the identity (\ref{rg}) is quite simple. 
Insert $G=G'+\Gamma$ in the expression of $A$ on the left hand side
 of (\ref{rg}) and decompose  the tree $T$
 into a family of connected subtrees
 $\{T_\alpha\}$ containing only
 $\Gamma$-lines and joined together by   $G'$-lines. Let
 $T'$ denote the tree which is obtained
from $T$ by contracting each tree  $T_\alpha$  to a point.
Now, let us fix  $T'$ and sum over the $\{T_\alpha\}'s$; this leads 
to the identity (\ref{rg}) with the ``renormalized'' vertices $w'$ 
given by Eq.\,\,(\ref{rg1}).
\vs{3mm}

We may  apply the identity (\ref{rg1}), starting with 
$w=w_0$, $G=G_0=G_1 +\Gamma_0$, $w'=w_1$, and then inductively 
to  $w=w_{n-1}$, $G=G_{n-1}=G_n +\Gamma_{n-1}$,
$w'=w_n$. It is easy to check that the $w_n's$ so defined
coincide with those constructed through Eq.\,\,(\ref{Wn+1}).  
Indeed, by the definition (\ref{W2}), 
$$W_1 (Y) = W_0(Y+\Gamma_0 W_1 (Y))\m.$$
Writing both sides in Fourier transform and expanding 
in a Taylor series, we obtain:
\qq
w_1 (Q;y) &=& \sum^\infty_{m=0} \sum_{ {\bf P}} w_1^{(m)}
(Q, {\bf P}) \prod^m_{{\ell} =
1} y (P_{\ell}) \non\\
&=& \sum^\infty_{m=0} \sum_{{\bf P}} w_0^{(m)} (Q,{\bf P})
\prod^m_{{\ell}=1} (y(P_{\ell}) + \Gamma_0\m w_1 (y) (P_{\ell}))
\label{tr4}
\qqq
Now expand the product over $\ell$ in the right hand side
replacing $w_1 (Q;y)$ in $\Gamma_0\m w_1(y)(P_\ell)$$= \sum_Q
\Gamma_0 (P_\ell,Q)\, w_1(Q;y)$ by the right hand side
of Eq.\,\,(\ref{tr4}). This leads, upon iteration, to
a sum over trees with a number of external lines to which a factor
$y(P_\ell)$ is attached. The trees
have an amplitude $A(T,\Gamma_0,w_0,{\bf Q})$, ${\bf Q}= (Q, {\bf P})$
and we can rewrite the right hand side of Eq.\,\,(\ref{tr4}),
using the definition (\ref{rg1}), as
$$
\sum^\infty_{m=0} \sum_{ {\bf P}} {w'_0}^{(m)}(Q,  {\bf P}) 
\prod^m_{\ell=1}y (P_\ell)\m.
$$
Comparison with the left hand side
of Eq.\,\,(\ref{tr4}) shows that $w_1=w'_0$ defined here
coincides with $w_1$ defined in Sect.\,\,2. The same arguments apply
inductively to $w_n$.
\vs{4mm}

Note that the transformation (\ref{rg1}) is rather easy to control
(in the sense that, if $w$ is small, $w'$ is also
small in a suitable norm)
and could be used to give
an  alternative proof of Proposition 3. 
Indeed, and this is the main difference between our approach
and the one of Eliasson, each transformation   (\ref{rg1})
involves small denominators (in $\Gamma_n$) on {\it  only one scale}.
To see intuitively how to use this fact, 
consider first all the trees having only vertices with
coordination number different from 2.
It is easy to see that these 
trees have a number of vertices with coordination number
one which is proportional to their total number of lines.
 Thus, one can control for those trees the small denominators 
on the lines, all of the same size,
by the exponential decay in $|q|$
of $w_n^{(0)}(Q)$ and the Diophantine condition (\ref{Dio}).
The next observation is that we may
  reduce ourselves to those trees
by resumming the contributions coming from the vertices 
with $m=1$ (i.e. coordination number 2). This is where the problem 
of repeated resonances appears in this formalism. We obtain a series 
of the form:
\qq
\sum^\infty_{k=0} (\Gamma_n w^{(1)}_n)^k\, \Gamma_n = (1-\Gamma_n
w^{(1)}_n)^{-1} \Gamma_n = \tilde
H\m \Gamma_n
\non
\qqq 
which has to be controlled using the Ward identities,
as in the proof of Proposition 3. Here we see again that this is the
subtle point of the proof.
\vs{4mm}

Finally, let us translate in the tree language some of the formulas
introduced in Sect.\,\,2.
Applying the resummation (\ref{rg}), (\ref{rg1}) to Eq.\,\,(\ref{sol}), 
we obtain:
 \qq
x(Q)=
{\sum_{Q_0}}^*\sum_{T\in\CT_1}G_{0}(Q,Q_0)\,\m 
A(T,G_1,w_1,Q_0)
\qqq
and, inductively,
 \qq
x(Q)=
{\sum_{Q_0}}^*\sum_{T\in\CT_1}
G_{0}(Q,Q_0)\,\m A(T,G_n,w_n,Q_0)
\qqq
which, upon the substitution $G_0=G_n + \Gamma_{<n}$,
may be rewritten as
\qq
x(Q)&=&
{\sum_{Q_0}}^*\sum_{T\in\CT_1}G_{n}(Q,Q_0)\,\m 
A(T,G_n,w_n,Q_0)\non\\
&+& {\sum_{Q_0}}^*\sum_{T\in\CT_1}
\Gamma_{<n}(Q,Q_0)\,\m A(T,G_n,w_n,Q_0)\m.
\label{tr3}
\qqq
This  corresponds to the decomposition
$$
X=F_n(Y)=Y + \Gamma_{<n} W_n(Y)\m,
$$
see Eq.\,\,(\ref{cum2}), with $Y$ satisfying $Y=G_n P W_n(Y)$. Indeed, 
the solution of the latter equation may be written, in the same way 
as the solution (\ref{sol}) of Eq.\,\,(\ref{fp1}), as a sum over 
the tree graphs. This gives the first term of on the right hand side 
of Eq.\,\,(\ref{tr3}). Moreover (compare with Eq.\,\,(\ref{wcl})),
$$
w_n (Q_0;y) =
\sum_{T\in\CT_1} A(T,G_n,w_n,Q_0)\m.
$$
This gives the second term on the right hand side 
of Eq.\,\,(\ref{tr3}).
\vskip 0.3cm

Note also that $X_n$, defined by the equality 
$X_n = \Gamma_{<n} W_0 (X_n)$, see Eq.\,\,(\ref{apeq}), 
satisfies the equality
\qq
x_n (Q)=
{\sum_{Q_0}}^*\sum_{T\in\CT_1}\Gamma_{<n}(Q,Q_0)\,\m
A(T,\Gamma_{<n},w_0,Q_0),
\label{soln}
\qqq
which may be derived from the equation $X_n=\Gamma_{<n} W_0(X_n)$
in the same way as we derived the identity (\ref{sol}).
Applying the relations (\ref{rg}), (\ref{rg1}) inductively, we 
we infer that
\qq
x_n (Q)
={\sum_{Q_0}}^*\Gamma_{<n}(Q,Q_0)\,\m w_n^{(0)}(Q_0)
\label{XN}
\qqq
which is equal to the term in the second sum
of Eq.\,\,(\ref{tr3}) corresponding to the tree
with only one vertex. It is also easy to see
why the other terms in Eq.\,\,(\ref{tr3}) are small and 
therefore why $x_n \to x$, as shown in the proof 
of Theorem 1: in all the other terms, the $w_n^{(0)}(p)$'s 
attached to vertices with coordination number one
are multiplied by some $G_n$, which forces the corresponding $p$ 
to be large (using the support properties of $G_n$,
the Diophantine property (\ref{Dio}) and the fact that
the sums here run over $p\neq 0$).
Then the exponential decay of $w_n^{(0)}(p)$ 
implies that the contribution of those trees is small.

\vspace*{8mm}

{\bf {Acknowledgments.}}
We thank L.H. Eliasson, G. Gallavotti and V. Mastropietro
for useful discussions.

\end{document}